\documentclass[aps,prl,twocolumn,superscriptaddress,floatfix]{revtex4-2}

\usepackage{amsmath,amssymb,graphicx,bm}
\usepackage[colorlinks=true,linkcolor=blue,citecolor=blue,urlcolor=blue]{hyperref}
\usepackage{tikz}
\usetikzlibrary{arrows.meta,decorations.pathmorphing,positioning}
\usepackage{orcidlink}

\begin{document}

\title{Constructing Non-Hermitian Theories with Tunable Exceptional Points and Controlled State Purification}

\author{Soumya Kanti Pal\,\orcidlink{0009-0008-7226-356X}}
\email{Corresponding Author: soumya.pal@tifr.res.in}

\author{Rupak Majumder\,\orcidlink{0009-0003-9275-8571}}
\email{rupak.majumder@tifr.res.in}

\author{Shamik Gupta\,\orcidlink{0000-0002-6080-4890}}
\email{shamik.gupta@theory.tifr.res.in}

\affiliation{Department of Theoretical Physics, Tata Institute of Fundamental Research, Homi Bhabha Road, Mumbai 400005, India}

\date{\today}

\begin{abstract}
Exceptional points (EP's) are a hallmark of non-Hermitian quantum systems. We show that momentum-space deformation provides a general design principle for creating and controlling EP's in quadratic many-body Hamiltonians. We identify universal criteria for the momentum sectors to host EP's and the corresponding critical deformation strengths, while revealing that a single momentum-sector EP induces quite remarkably an exponential proliferation of many-body eigenvector coalescences. We further establish EP's as a universal mechanism for purifying arbitrary mixed quantum states, uncovering distinct purification regimes and a fundamental odd-even system-size dichotomy in the thermodynamic limit. Our framework also provides a systematic reverse-engineering protocol for generating short- and long-range, reciprocal and nonreciprocal non-Hermitian quantum matter, together with an explicit Lindblad embedding. These results thus establish momentum-space deformation as a unified route to exceptional-point engineering and controlled design of many-body non-Hermitian quantum systems.
\end{abstract}

\maketitle

\textit{Introduction:--} Since the discovery that non-Hermitian Hamiltonians can possess entirely real spectra under suitable symmetry constraints~\cite{Bender1998,Bender2007}, non-Hermitian quantum mechanics has emerged as a fundamentally new framework for describing open and driven systems. A hallmark of these systems is the exceptional point (EP), a non-Hermitian spectral singularity at which both eigenvalues and eigenvectors simultaneously coalesce~\cite{Kato1995,Heiss2012}. Unlike ordinary degeneracies in Hermitian systems, EP's render the Hamiltonian non-diagonalizable, giving rise to nontrivial topology in parameter space~\cite{Dembowski2001,Berry2004} and novel quantum dynamics~\cite{Uzdin2011,Dora2019,Zhang2020,Wu2020,Xiao2021,Gopalakrishnan2021,Weidemann2022,Sim2023,Bai2024, Agarwal2024,Sim2025}. Over the past decade, EP's have become central to modern non-Hermitian physics, underpinning phenomena ranging from generalized bulk-boundary correspondence and topological phases~\cite{Kunst2018,Kawabata2019,Bergholtz2021,Ding2022_1} to chiral state transfer~\cite{Lefebvre2009,Berry2011,Khandelwal2024,Gao2025}, enhanced sensing~\cite{Wiersig2014,Chen2017,Hokmabadi2019,Yu2020}, transport~\cite{Saha2023, Saha2023_1, Saha2025}, and modified linear responses~\cite{Hashemi2022,Geier2022}, which have no counterpart in Hermitian quantum mechanics~\cite{Doppler2016,Xu2016,Chen2017,Hodaei2017,Ashida2020}.

On the experimental side, rapid advances have realized EP's in diverse platforms, including microwave cavities~\cite{Dembowski2001}, photonic systems~\cite{Ding2016,Thomes2024}, thermal atoms~\cite{Zhang2024}, superconducting qubits~\cite{Chen2021,Chen2022}, diamond~\cite{Wu2025}, electrical circuits~\cite{Helbig2020}, non-Markovian quantum systems~\cite{Lin2025,Zhang2025}, and quantum heat engines~\cite{Zhang2022,Bu2023}. These developments have established EPs as experimentally accessible resources for quantum technologies, motivating the search for general principles to engineer and control them.

A general strategy for exploring exceptional-point physics is to investigate specific non-Hermitian Hamiltonians and characterize their spectral and dynamical properties. Together with symmetry and topological classifications~\cite{Ashida2020,Kawabata2019}, these studies have established a broad understanding of non-Hermitian phases~\cite{Kawabata2019}. An important complementary question is: Can one systematically engineer EP structures in extensive classes of many-body Hamiltonians and, if so, what are their consequences for many-body physics? 

In this Letter, we introduce momentum-space deformations as a general framework for engineering non-Hermitian many-body systems with controllable exceptional-point (EP) structures, and provide a direct bridge with many-body physics. Quadratic Bogoliubov--de Gennes (BdG) Hamiltonians, whose momentum-space representation decomposes into independent blocks, provide a natural setting that connects momentum-space deformations, their real-space counterparts, and the resulting many-body physics. We derive universal, microscopic-independent criteria determining both the momentum sectors hosting EP's and their critical deformation strengths, and prove that a single momentum-sector EP induces an exponential proliferation of pairwise eigenvector coalescences throughout the many-body spectrum. Whereas generic non-Hermitian perturbations restrict EP's to a subset of momentum sectors, $\mathcal{PT}$ symmetry enables every block to host an EP, completely characterizing the associated many-body $\mathcal{PT}$-symmetry-breaking transition. We further establish EP's as a universal mechanism for purifying arbitrary mixed quantum states, uncovering three purification regimes together with a striking odd-even dichotomy: complete purification is impossible for odd system sizes but remains attainable for even ones, yielding fundamentally inequivalent thermodynamic limits.

Moreover, momentum-space deformations offer a systematic reverse-engineering protocol for generating short- and long-range, reciprocal and nonreciprocal non-Hermitian Hamiltonians. We also provide an explicit Lindblad embedding that realizes arbitrary quadratic non-Hermitian Hamiltonians in open quantum systems. Together, these results establish momentum-space deformation as a general framework for exceptional-point engineering in quadratic many-body systems.

\textit{Free fermionic Hamiltonians:---} Any generic quadratic Hamiltonian on a periodic lattice of $L$ spinless fermions can be written as~\cite{Bogoliubov1958, deGennes2018, Mbeng2024} 
\begin{align}
    \hat{H} =  \Psi^\dagger \mathcal{H}_{\text{BdG}} \Psi 
    \label{eq: H real space}
\end{align}
in terms of the Nambu spinor $\Psi^\dagger\equiv \begin{pmatrix}
    \hat{c}_1^\dagger & \hat{c}_2^\dagger & \ldots &  \hat{c}_L^\dagger & \hat{c}_1 &  \ldots \hat{c}_L
\end{pmatrix}$, with $\hat{c}_j, \hat{c}_j^\dagger$ being the fermionic annihilation and creation operators at site $j$. The Bogoliubov De-Gennes (BdG) matrix $\mathcal{H}_{\text{BdG}}$ encodes the hopping and pairing amplitudes. Such systems are most conveniently studied via Fourier transformation to the momentum basis:  $\hat{c}_j = (1/\sqrt{L})\sum_{k}e^{ijk} \hat{c}_k $, where the momentum modes $k$ are restricted to the first Brillouin zone. In this basis, the Hamiltonian retains the same quadratic form, up to a constant: $\hat{H} =  \widetilde{\Psi}^\dagger \widetilde{\mathcal{H}}_{\text{BdG}} \widetilde{\Psi}$, with $\widetilde{\Psi}^\dagger$ being the Fourier-transformed Nambu spinor. For systems with translational symmetry, $\widetilde{\mathcal{H}}_{\text{BdG}}$ block-decomposes into Bloch Hamiltonian $\mathcal{H}_k$:
\begin{align}
    \hat{H} = \sum_{k \in \mathrm{RBZ}} \Psi_k^\dagger \mathcal{H}_k \Psi_k, \label{eq: H momentum basis}
\end{align}
with $\mathcal{H}_k =  \mathcal{H}_k^\dagger$, $\Psi_k$'s the Nambu-Gorkov spinors, and RBZ the reduced Brillouin zone  \footnote{The dimension of each $\mathcal{H}_k$ is determined by the size of the unit cell associated with the underlying translational symmetry. For instance, the transverse-field Ising model with a staggered transverse field, $h\sum_i (-1)^i \sigma_i^x$, has a two-site unit cell, resulting in a $4\times4$ Bloch Hamiltonian $\mathcal{H}_k$}.

For a homogeneous system, $\mathcal{H}_k$ is a $(2 \times 2)$ matrix, and we may take $\Psi_k^\dagger = \begin{pmatrix}
    c_k^\dagger & i c_{-k}
\end{pmatrix}$, with $k\in \mathrm{RBZ} \equiv \{k:[0,\pi]\}$. The modes $k=0,\pi$ are self-conjugate, and the corresponding $\mathcal{H}_k$'s are numbers. For the remaining $k$'s, $\mathcal{H}_k$ has the generic form $\mathcal{H}_k = \begin{pmatrix}
    A(k) & B (k )\\
    C(k) & D(k)
\end{pmatrix}$,
where all of $A, B, C, D$ are $2\pi$-periodic functions of $k$, and Hermiticity demands $A$ and $D$ to be real and $C=B^{*}$.

Now, we move on to construct generic non-Hermitian BdG theories with $\mathcal{H}_k^{\mathrm{nH}}$, such that $\mathcal{H}_k^{\mathrm{nH}} \neq (\mathcal{H}_k^{\mathrm{nH}})^\dagger$. One can always decompose an arbitrary $\mathcal{H}_k^{\mathrm{nH}}$ in terms of two Hermitian matrices as $\mathcal{H}_k^{\mathrm{nH}} = H^{(1)}-iH^{(2)}$, where $H^{(1)} = (\mathcal{H}_k^{\mathrm{nH}}+ \mathcal{H}_k^{\mathrm{nH} \dagger})/2$ and $H^{(2)} = i(\mathcal{H}_k^{\mathrm{nH}}- \mathcal{H}_k^{\mathrm{nH} \dagger})/2$. This decomposition may be viewed as a non-Hermitian deformation of the Hermitian Hamiltonian $H^{(1)}$ by $H^{(2)}$. Such non-Hermitian Hamiltonians often host EP's. 

\begin{figure}
    \centering
    \includegraphics[width=0.8\linewidth]{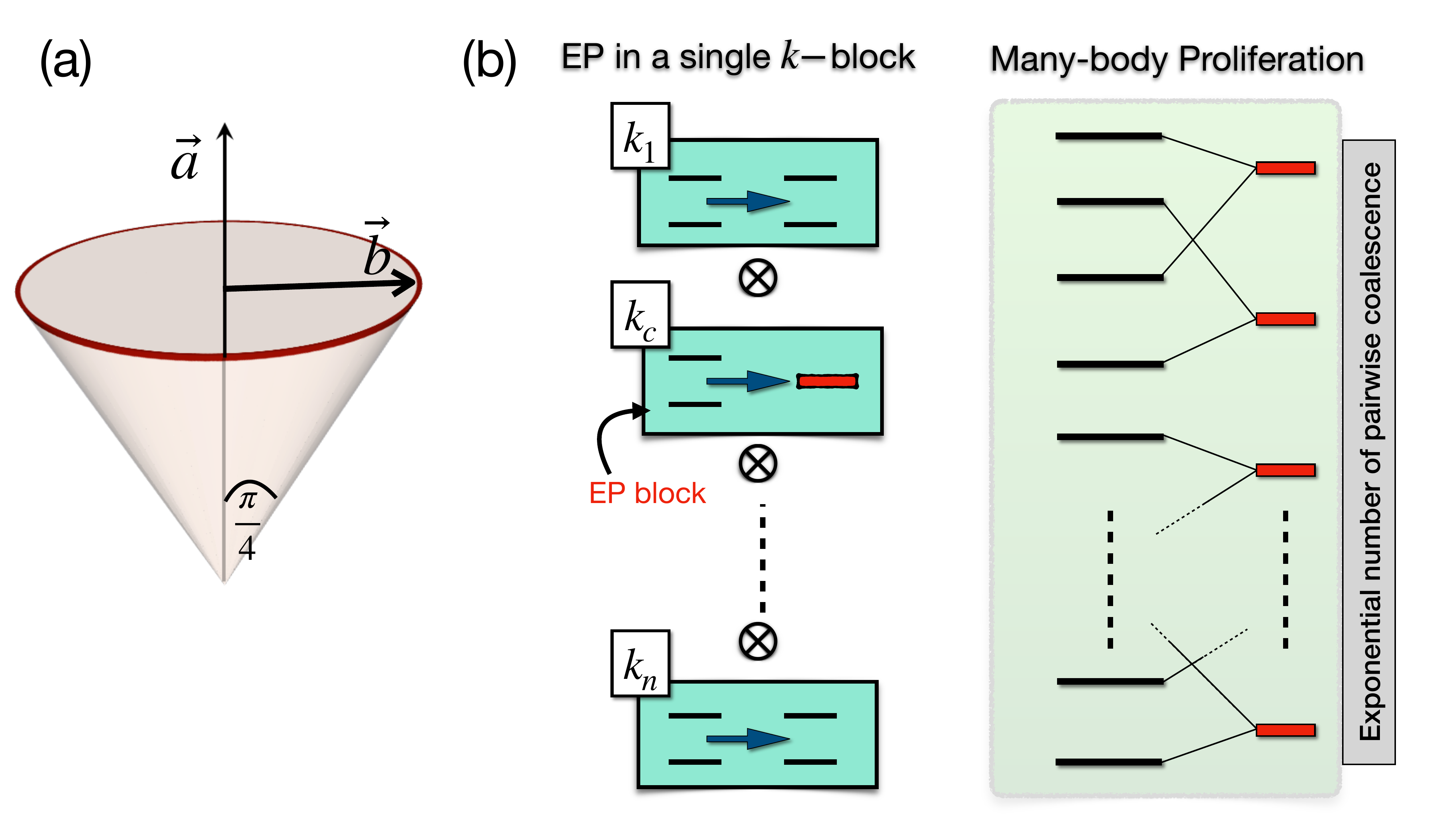}
    \caption{(a): Geometric representation of EP conditions~\eqref{eq:EP-Condition}. (b): Exponential proliferation of many-body eigenvector coalescences induced by an EP in a single momentum block.}
    \label{fig: fig 1}
\end{figure}

We next discuss the precise conditions for the existence of EP's. Considering $H^{(1)}=\mathcal{H}_k$, EP existence depends crucially on the choice of the deformation $H^{(2)}$. Given that any $(2\times 2)$ matrix can be expressed in terms of Pauli matrices $\{\hat{\sigma}^\alpha\}_{\alpha=x,y,z}$  and the identity $\mathbb{I}_2$, we use the parametrization $H^{(1)} = a_0 \mathbb{I} + \vec{a}\cdot \vec{\sigma}$ and $H^{(2)} = b_0 \mathbb{I} + \vec{b}\cdot \vec{\sigma}$, where $a_0 ,b_0, \vec a, \vec b$ are $2\pi$-periodic in $k$. Consequently,  $\mathcal{H}_k^{\mathrm{nH}}= (a_0-ib_0)\mathbb{I}+(\vec{a}-i\vec{b})\cdot \vec{\sigma}$ has eigenvalues $ E^{(\pm)}_k = (a_0-ib_0)\pm\sqrt{a^2 - b^2-i2\vec{a}\cdot \vec{b}}$, where $a=|\vec{a}|, b=|\vec{b}|$. Obtaining the corresponding eigenvectors and imposing the EP condition, namely, the simultaneous coalescence of both eigenvalues and eigenvectors, yields the following conditions for $\mathcal{H}_k^{\mathrm{nH}}$ to host an EP: 
\begin{align} \label{eq:EP-Condition}
\mathrm{(i)}~\vec{a}\cdot \vec{b} = 0,~~\mathrm{(ii)}~a = b 
\end{align}
with $\vec{a} \neq \vec{b}$ and arbitrary $a_0$, $b_0$ \cite{supplement}. Thus, given $\vec{a}$, the vector $\vec{b}$ has to lie on the perpendicular plane and is of the same length as $\vec{a}$. As in Fig.~\ref{fig: fig 1}(a), this defines a cone with $\vec{a}$ as the axis of symmetry and half angle of $\pi/4$ radians, such that every point on its surface is an EP.

Without loss of generality, we make the choice $\vec{b}^\dagger = \begin{pmatrix}
    \gamma_xF_x(k) & \gamma_yF_y(k) & \gamma_zF_z(k)
\end{pmatrix} $, where the real quantity $\gamma_\alpha F_\alpha(k)$ denotes deformation of strength $\gamma_\alpha$ along direction $\alpha$. By considering deformations along individual directions, we now systematically determine the absence or existence of EP's, together with their locations in the $(k,\gamma_{\alpha})$ plane. Rewriting $\mathcal H_k$ as
\begin{equation}
    \mathcal{H}_k = \left[\frac{A+D}{2}\right]\mathbb{I} + \mathrm{Re}\left(B\right)\sigma^x-\mathrm{Im}\left(B\right)\sigma^y+\left[\frac{A-D}{2}\right]\sigma^z, \label{eq: Hk in a}
\end{equation}
EP condition (i) then gives the particular momentum block $k_c \,(\neq 0,\pi)$ at which EP occurs, while condition (ii) gives the corresponding critical strength $\gamma_{\alpha}^{(c)}$ as
\begin{align}
    \gamma_{\alpha}^{(c)} = \left[2F_\alpha(k_c)\right]^{-1}\sqrt{4\left|B(k_c)\right|^2+\left[A(k_c)-D(k_c)\right]^2}.\label{eq: gamma c}
\end{align}
On the other hand, the EP block $k_c$ satisfies: $\mathrm{Re}\left(B(k_c)\right) = 0$ for $\alpha = x$,   $A(k_c)=D(k_c)$ for $\alpha =z$, and  $\mathrm{Im}\left(B(k_c)\right) = 0$ for $\alpha = y$ \footnote{Condition (i) also gives $F_\alpha(k_c)=0$ which leads to diverging $\gamma_c$, hence discarded.}. Remarkably,  $k_c$ is determined solely by the intrinsic structure of \(\mathcal{H}_k\), independent of the specific choice of the non-Hermitian deformation; the latter only controls the value of \(\gamma_\alpha^{(c)}\). The simultaneity of the two EP conditions precludes the possibility of an arbitrary $k_c$ by merely tuning $\gamma_\alpha$. However, if $B(k)$ is either purely real or imaginary, the criterion determining $k_c$ is trivially satisfied for $\alpha =y$ and $\alpha = x$, respectively. This provides a powerful mechanism for engineering exceptional points in any momentum block simply by tuning the corresponding $\gamma_\alpha$.

We now show that the aforementioned mechanism is trivially satisfied for a Hamiltonian with $\mathcal{PT}$-symmetry. In real space, one can easily show that the parity operator $\mathcal{P}$ acts on fermionic operators as $\hat{\mathcal{P}}\hat c_i\hat{\mathcal{P}}^{-1}
= (-1)^{\hat N_f}\hat c_{L+1-i}$ and $\hat{\mathcal{P}}\hat c_i^\dagger\hat{\mathcal{P}}^{-1}
= -(-1)^{\hat N_f}\hat c_{L+1-i}^\dagger$, with $\hat N_f$ the fermionic number operator. The time-reversal operator $\mathcal{T}$ is anti-unitary and satisfies $\hat{\mathcal{T}}\, i\, \hat{\mathcal{T}}^{-1}=-i$ along with $\hat{\mathcal{T}}\hat c_i\hat{\mathcal{T}}^{-1}
= \hat c_{i}$ and $\hat{\mathcal{T}}\hat c_i^\dagger\hat{\mathcal{T}}^{-1}
= \hat c_{i}^\dagger$. It follows that $\mathcal{PT}$ symmetry constrains the non-Hermitian Hamiltonian such that $\vec{a}$  has no $y$ component, whereas $\vec{b}$ is restricted to have only the $y$ component, leading to $\mathcal{H}_k^{\mathrm{nH}} = a_0\mathbb{I} + a_x\sigma^x - i b_y\sigma^y + a_z\sigma^z$~\cite{supplement}. Evidently, the EP condition (i) is satisfied. Consequently, every momentum block $k$ of a $\mathcal{PT}$-symmetric $\hat{H}^{\mathrm{nH}}$ hosts an EP, provided the non-Hermitian deformation strength is tuned to its corresponding critical value, $\gamma_y=\gamma_y^{(c)}(k)$. On either side of the EP, (i) when $a<b$, the eigenvalues $E_k^{(\pm)}$ are purely real, corresponding to $\mathcal{PT}$-symmetry-preserved (SP) phase, or, (ii) when $a>b$, the eigenvalues are purely imaginary, corresponding to $\mathcal{PT}$-symmetry-broken (SB) phase.

\textit{An illustrative example:--} Consider the transverse field Ising Chain (TFIC), for which the Hamiltonian in the real space spinless-fermionic representation reads as
\begin{align}
    \hat{H} &= -J \sum_{i=1}^{L}\Big[\hat{c}_i^\dagger \big(\hat{c}_{i+1}^\dagger+\hat{c}_{i+1}\big)+\mathrm{h.c.}\Big]+2h\sum_{i=1}^{L}c_i^\dagger c_i, \label{eq: TFIC}
\end{align}
where $\hat{c}_{L+1} = \mathcal{P}_f\hat{c}_1$ with $\mathcal{P}_f$ being the fermionic parity operator. In the momentum basis, this Hamiltonian takes the form~\eqref{eq: H momentum basis} with $A(k) = -D(k)=2(h-J\cos{k})$ and $B(k) = C(k)= 2J \sin{k}$, where $ k  =  (2m+1)\pi/L $ for even parity and $k = \pi, \pm 2m\pi/L$ for odd parity and $m\in \{0,1,\cdots,L/2-1\}$, see standard references, e.g.,~\cite{Mbeng2024, SachdevBook}~and~\cite{supplement}. Considering deformation along $x$, the EP condition (i), namely, $\mathrm{Re}\left(B(k_c)\right)=\sin{k_c}= 0 $ requires $k_c = 0,\pi$ that are not valid momentum modes to host EP. For $z$-deformation, the corresponding condition $A(k_c) = D(k_c)$ gives $k_c = \cos^{-1}\left(h/J\right)$ (provided $|h|\leq J$), while Eq.~\eqref{eq: gamma c} gives $\gamma_z^{(c)} = \left[2F_\alpha(k_c)\right]^{-1}\sqrt{J^2-h^2}$. Fixing $J$ and tuning $h$ from higher to lower values, no EP appears for any values of $\gamma$ until $h$ hits the value $J \cos{[(L-1)\pi/L]}$ and $\gamma=\gamma_z^{(c)}$, when the system shows its first EP in the momentum block $k_c = (L-1)\pi/L$. Decreasing $h$ further to the value $J \cos{[(L-2)\pi/L]}$ leads to the next EP in the contiguous momentum block. This continues until $h$ crosses $-J$, when the system does not show any further EP. Evidently, as $L\to \infty$, the system hosts EP at all $|h|< J$ values. For $y$-deformation, the EP condition (i): $\mathrm{Im}\left(B(k_c)\right)=0$ is trivially satisfied, implying we can induce EP in any $k$ block by tuning $\gamma_y$ to $\gamma_y^{(c)} = 2\left[F_y(k_c)\right]^{-1}\sqrt{h^2+J^2-2Jh\cos{k_c}}$. To summarize, $x$-deformation results in no EP for any values of $(h,J)$, $z$-deformation induces EP in specific momentum blocks for $|h|<J$, and $y$-deformation induces EP in all momentum blocks for all $(h,J)$.

\textit{Many-body proliferation of a single-block EP:--}  It is instructive now to examine how the emergence of EP's in $\mathcal{H}^\mathrm{nH}_k$ manifests in the eigenvalues and eigenvectors of $\Psi_k^\dagger \mathcal{H}^\mathrm{nH}_k \Psi_k$, and consequently of the many-body spectrum of the parent Hamiltonian $\hat{H}^\mathrm{nH}$. The two eigenvalues $E_k^{(\pm)}$ of $\mathcal{H}^\mathrm{nH}_k$ will also be the eigenvalues of $\Psi_k^\dagger \mathcal{H}^\mathrm{nH}_k \Psi_k$ with the corresponding eigenstates
\begin{equation}
    |\psi_k^{(\pm)}\rangle = \mathcal{N}\Big[i C(k)|0\rangle + \big[E_k^{(\pm)}-D(k)\big]c^\dagger_kc^\dagger_{-k}|0\rangle\Big],
\end{equation}
where $|0\rangle$ is the vacuum state and $\mathcal{N}$ is the normalization constant~\cite{Mbeng2024}. When $\mathcal{H}^\mathrm{nH}_k$ hosts an EP, the eigenvalues $E_k^{(\pm)}$ coalesce, and so do the corresponding eigenvectors $|\psi_k^{(+)}\rangle$ and $ |\psi_k^{(-)}\rangle$. Consequently, the operator $\Psi_k^\dagger \mathcal{H}^\mathrm{nH}_k \Psi_k$ also hosts an EP. Using $[\Psi_k^\dagger \mathcal{H}^\mathrm{nH}_k \Psi_k, \Psi_{k'}^\dagger \mathcal{H}^\mathrm{nH}_{k'} \Psi_{k'}]=0~\forall~k,k'$, the energy eigenvalues of  $\hat{H}^{\mathrm{nH}}$ reads as $\mathbb{E}_{\{s_k\}}= \sum_{k}E^{(s_k)}_k$, with $s_k = \pm1$, with the corresponding many-body eigenstate $|\Phi_{\{s_k\}}\rangle = \bigotimes_k |\psi^{(s_k)}_k \rangle$. All possible combinations of the string $\{s_k\}$ generate the entire many-body spectrum. When the block $k = k_c$ hosts an EP, we will have $\mathbb{E}_{\{\cdots s_{k<k_c},+1,s_{k>k_c},\cdots\}}=\mathbb{E}_{\{\cdots s_{k<k_c},-1,s_{k>k_c},\cdots\}}$ for all possible $\{s_{k\neq k_c}\}$ combinations, leading to the coalescence of the corresponding eigenvectors, $|\Phi_{\{\cdots s_{k<k_c},+1,s_{k>k_c},\cdots\}}\rangle = |\Phi_{\{\cdots s_{k<k_c},-1,s_{k>k_c},\cdots\}}\rangle$. The number of such pairwise coalescences is $2^{L/2-1}$ in the even parity sector, and $2^{L/2-2}$ in the odd parity sector~\footnote{We have taken $L$ to be even}. Thus, an EP in a single momentum block induces an exponential proliferation of eigenvector coalescences in the many-body Hilbert space.  A schematic of this proliferation is shown in Fig.~\ref{fig: fig 1}(b).

\textit{Dynamical signature of EP and state purification:--}  We now investigate the dynamical signatures of EP's, first at the level of a single momentum block and subsequently in the many-body Hilbert space. To this end, we show that evolution under an $(n\times n)$ non-Hermitian matrix with EP of order $n$ can purify any arbitrary mixed state (Appendix A). Such mixed-state purification~\cite{Childs2025} is a key resource for information processing, including error correction~\cite{Shor1995,Steane1996,Bennett1996}, cooling~\cite{Schulman2005,Ticozzi2014}, and resource distillation~\cite{Bennett1996_1,Bravyi2005,Fang2020}, making purity a natural dynamical diagnostic of EP's with direct relevance to quantum technologies.

Consider a generic initial state characterized by the  density matrix $\rho(t=0) = \otimes_{k\ge0} \rho_k(0)$, with $\rho_k(0)=(1/2)(\mathbb{I}+\vec{p}\cdot\vec{\sigma})$. Under the evolution of  $\mathcal{H}_k^{\text{nH}}$, the density matrix at time $t$ becomes $\rho_k(t)=(1/2)\left[p_0(t)\mathbb{I}+\vec{p}(t)\cdot\vec{\sigma}\right]$,
where $p_0(t),~\vec{p}(t)$ can be obtained analytically. The corresponding normalized purity at time $t$ is $P_k(t)= \big[1 + |\vec{p}(t)|^2/p^2_0(t)\big]/2$, and that for the full-Hilbert space is $P(t) = \prod_{k\ge 0} P_k(t)$. Complete purification ($P(t)=1$) has been demonstrated earlier to be possible in specific non-Hermitian Hamiltonians that are $\mathcal{PT}$-symmetric~\cite{Gopalakrishnan2021}. Here, we elevate the attainment of purification to a design principle: Our framework identifies a broad class of engineered $\mathcal{PT}$-symmetric many-body Hamiltonians that guarantee complete purification of arbitrary mixed states and, more generally, enables programmable control of the asymptotic many-body purity by tuning the non-Hermitian deformation strength. Specifically, starting from the maximally mixed state with $\vec{p}= \mathbf{0}$, the purity of the momentum block $k$ at time $t$ equals
\begin{equation}
    P_k(t) = \frac{1}{2}+\frac{2b^2\sin^2{(\omega_k t)}\big[a^2-b^2\cos^2{(\omega_k t)}\big]}{\big[a^2-b^2\cos{(2\omega_k t)}\big]^2}, \label{eq: purity} 
\end{equation}
where  $\omega_k = \sqrt{a^2-b^2}$. In the regime where $k$ block Hamiltonian is in the SP phase with $a>b$, we have $\omega_k\in \mathbb{R}$, and Eq.~\eqref{eq: purity} implies that the purity oscillates in time. In contrast, in the SB phase with $a<b$, $\omega_k$ becomes purely imaginary, leading to $P_k(t\to \infty) = 1$. At the common boundary of the two phases, which corresponds to an EP, we have $\omega_k = 0$, leading to $P_k(t\to \infty)= 1$. Generalizing this, we are able to show that
in the SB phase and on its boundary, any mixed state gets completely purified at long times (Appendix B).

Going beyond individual blocks, the many-body spectrum will be in the SP phase when all $\omega_k$'s are real, and in the  SB phase when they are purely imaginary. As unveiled earlier, every momentum block $k$ of the $\mathcal{PT}$-symmetric $\hat{H}^{\mathrm{nH}}$ hosts an exceptional point at $\gamma_y=\gamma_y^{(c)}(k)$ [Eq.~\eqref{eq: gamma c}], constituting the boundary between the SP and SB phases. This gives rise to two many-body thresholds, $\gamma_\mathrm{c}^{(1)}\equiv\min_{k\in \mathrm{RBZ}}\gamma_y^{(c)}(k)$ and $\gamma_\mathrm{c}^{(2)}\equiv\max_{k \in \mathrm{RBZ}}\gamma_y^{(c)}(k)$, at which a single momentum block hosts an exceptional point, and the remaining blocks are in either an SP phase at $\gamma_y = \gamma_c^{(1)}$ or an SB phase at $\gamma_y = \gamma_c^{(2)}$. Consequently, for $\gamma_y<\gamma_\mathrm{c}^{(1)}$, all momentum blocks remain in the SP phase, and for an initial mixed state, the many-body purity $P(t)$ exhibits persistent oscillations near its initial value. In the intermediate regime, $\gamma_\mathrm{c}^{(1)}<\gamma_y<\gamma_\mathrm{c}^{(2)}$, only a subset of momentum blocks is in the SB phase, and $P(t)$ relaxes to persistent oscillations around a higher value, signaling partial purification. For $\gamma_y>\gamma_\mathrm{c}^{(2)}$, every momentum block and thus the entire system is in SB phase. Hence, the many-body purity saturates to maximum allowed value without oscillations, implying complete purification of any initial mixed state. Thus, depending on $\gamma_y$, the many-body purity exhibits three distinct purification regimes.

\begin{figure}
    \centering
    \includegraphics[width=0.8\linewidth]{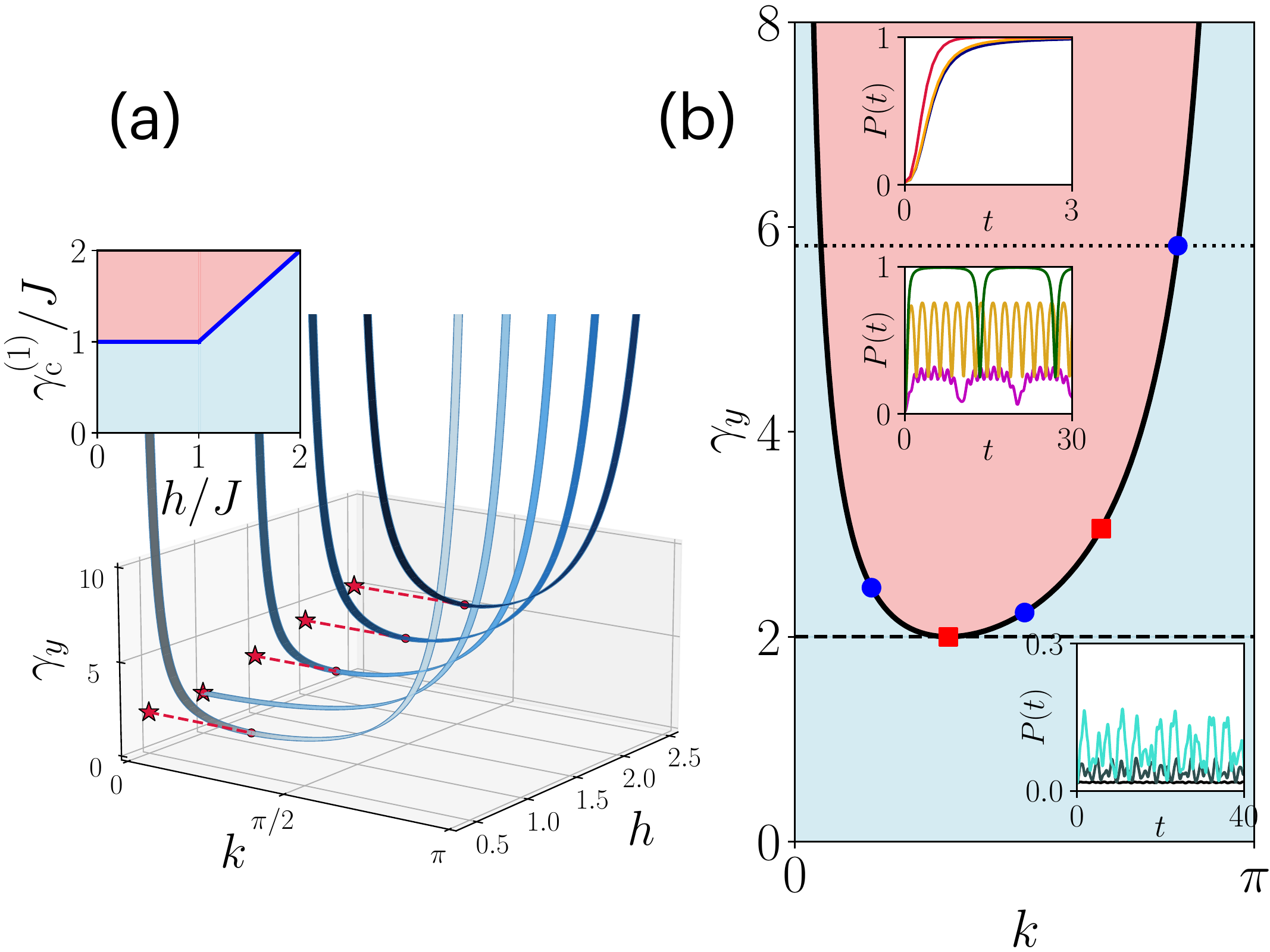}
    \caption{For the non-Hermitian Hamiltonian $H=H^{(1)}-iH^{(2)}$, with $H^{(1)}$ in Eq.~\eqref{eq: TFIC} and the $y$-deformation
$H^{(2)}=\gamma_y\sin k\,\sigma^y$, every allowed momentum block hosts an EP at $k_c=k\in\mathrm{RBZ}$. (a) Critical deformation strength $\gamma_y^{(c)}$ in Eq.~\eqref{eq: gamma c} as a function of $(h,k,\gamma_y)$. The stars mark the minimum threshold $\gamma_\mathrm{c}^{(1)}\equiv\min_{k\in\mathrm{RBZ}}\gamma_y^{(c)}(k)$, while the inset shows $\gamma_\mathrm{c}^{(1)}/J$ versus $h/J$. (b) For $h=0.5$, $J=1$, the solid black curve shows $\gamma_y^{(c)}(k)$ in the thermodynamic limit, with the dashed horizontal line indicating $\gamma_\mathrm{c}^{(1)}$. For a finite system with $L=6$, the red (blue) circles denote the odd-parity (even-parity) momentum blocks. Although the upper threshold $\gamma_\mathrm{c}^{(2)}\equiv\max_{k\in\mathrm{RBZ}}\gamma_y^{(c)}(k)$ diverges in the thermodynamic limit, its finite-size value is indicated by the dotted horizontal line. Going from top to bottom, the three subpanels show the time evolution of the many-body purity, starting from the maximally mixed state, in the regimes (i) $\gamma_y (=0.5, 1.5,1.99)<\gamma_\mathrm{c}^{(1)}$, (ii) $\gamma_\mathrm{c}^{(1)}<\gamma_y(=2.4, 5.0,5.8)<\gamma_\mathrm{c}^{(2)}$, and (iii) $\gamma_y(=5.82,6.0,8.0)>\gamma_\mathrm{c}^{(2)}$. }
    \label{fig: fig 2}
\end{figure}

To illustrate these features, we consider the Hamiltonian in Eq.~\eqref{eq: TFIC} with the non-Hermitian $k$-block deformation $H^{(2)}=\gamma_y\sin k ~\sigma^y$. From Eq.~\eqref{eq: gamma c}, the critical deformation strength is given by $\gamma_y^{(c)}(k)=2\big[\sin k\big]^{-1}\sqrt{h^2+J^2-2Jh\cos k}$. Its dependence on $k$ and $h$ is shown in Fig.~\ref{fig: fig 2}(a). The minimum value $\gamma_c^{(1)}$ equals $J$ for $h\le J$ and increases linearly with $h$ for $h>J$, Fig.~\ref{fig: fig 2}(a)(inset). On the other hand, the maximum value $\gamma_c^{(2)}$ diverges in the thermodynamic limit $L \to \infty$, see Fig.~\ref{fig: fig 2}(b). The momentum blocks lying in the red region are in the SB phase and undergo complete purification, whereas the blocks lying in the blue region are in the SP phase and their purity exhibits persistent oscillations near its initial value, Fig.~\ref{fig: fig 2}(b).

In contrast to the thermodynamic limit, the threshold $\gamma_c^{(2)}$ is finite for finite $L$, rendering complete purification of a many-body state possible even for a finite $\gamma_y$. Here, however, the value of this threshold depends crucially on whether $L$ is even or odd and on the fermionic parity, which determine the discrete momentum values $\{k\}$ and if unpaired modes $k=0,\pi$ are present. For even $L$, the even-parity sector contains only paired modes, allowing full state purification $P(t\to\infty)=1$ for $\gamma_y > \gamma_c^{(2)}$. Conversely, the odd-parity sector contains two unpaired modes ($k=0,\pi$), capping the purity at $P(t\to \infty)=0.25$. For odd $L$, both parity sectors contain a single unpaired mode, $k=\pi$ for even and $k=0$ for odd parity, yielding $P(t\to\infty)=0.5$ when $\gamma_y > \gamma_c^{(2)}$. This establishes a fundamental odd-even dichotomy: complete purification is unattainable for odd system sizes but remains possible for even ones, rendering their thermodynamic limits inequivalent. Results corresponding to even $L$ are presented in Fig.~\ref{fig: fig 2}, while odd-$L$ results are in Appendix C. Note that thermodynamic divergence of $\gamma_c^{(2)}$ can be avoided by choosing $F(k)=1$ (Appendix D), which generates a long-range real-space Hamiltonian, as we show next. By contrast, $F(k)=\sin k$ considered above yields a short-range model.

\textit{Constructing short-range, long-range and nonreciprocal theories---} Here, we demonstrate how momentum-space deformations offer a powerful framework for tailoring the real-space structure of non-Hermitian Hamiltonians. For our choice of Nambu-Gorkov spinor $\Psi_k$, the operator $\hat S^\alpha(k)\equiv \Psi_k^\dagger \sigma^\alpha \Psi_k$ is antisymmetric in $k$ for $\alpha  = x,y$, and is symmetric for $\alpha=z$. Consider a deformation along $\alpha$, so that Eq.~\eqref{eq: H momentum basis} reads as $\hat{H}^{\text{nH}} = \hat{H} + \delta \hat{H}_{\alpha} $, where $\delta \hat{H}_{\alpha} = -i\gamma_{\alpha} \sum_{k\ge 0} F_{\alpha} (k) \hat{S}^{\alpha}(k)$, whose real-space representation is obtained via inverse Fourier transformation involving momentum contributions over the full Brillouin zone (BZ). Antisymmetry of $\hat S^\alpha(k)$ results in two distinct classes dictated by the parity of $F_{\alpha}(k)$: (a) $F_{\alpha}(k)=F_{\alpha}(-k)$ gives $\delta \hat H_{\alpha} = (i\gamma_\alpha/2) \sum_{k\in \text{BZ}}F_{\alpha} (k) \hat{S}^{\alpha}$; on Fourier transforming, we get $\hat{H}^{\text{nH}}$ comprising $\hat{H}$ in Eq.~\eqref{eq: H real space} augmented by $\delta \hat H_{\alpha}$ that involves short-range real-space couplings $\sim \hat{c}_i \hat{c}_{i+r}$ with $r\ll L$;
(b) $F_{\alpha}(k)=-F_{\alpha}(-k)$ gives $\delta \hat H_{\alpha} = (i\gamma_\alpha/2) \sum_{k\in \text{BZ}} \mathrm{sgn}(k)F_{\alpha} (k) \hat{S}^{\alpha}$. Since the Fourier transform of ${\rm sgn}(k)$ decays algebraically as $\sim1/r$, $\hat{H}^{\text{nH}}$ in real space has intrinsically long-range couplings, e.g., class (a) with $F_y(k) =\sin{k}$ gives $\delta \hat H_y = (\gamma_y/2) \sum_{j=1}^L \left[\hat c_j^\dagger \hat c_{j+1}^\dagger + \hat c_j \hat c_{j+1} \right] $, while (b) with $F_y(k)=1$ gives $\delta \hat H_y =(2\gamma_y/\pi) \sum_j \sum_{r>0, ~r\in \mathrm{Odd}} \left[ \hat{c}_j^\dagger\hat{c}_{j+r}^\dagger + \hat{c}_j \hat{c}_{j+r}  \right]/r $. Other representative examples are detailed in~\cite{supplement}. 

Quite remarkably, considering  simultaneous deformation along both $x$, $y$, i.e., $\delta \hat{H} = -i\gamma_{x} \sum_{k\ge 0} F_{x} (k) \hat{S}^{x}(k)-i\gamma_{y} \sum_{k\ge 0} F_{y} (k) \hat{S}^{y}(k)$, naturally results in real-space Hamiltonians with nonreciprocal couplings. For example, $F_x(k) = F_y(k) = \sin{k}$ yields
\begin{equation}
    \delta \hat{H} = \frac{1}{2}\sum_{j=1}^L \left[(\gamma_y+i\gamma_x)\hat c_j^\dagger \hat c_{j+1}^\dagger + (\gamma_y-i\gamma_x)\hat c_j \hat c_{j+1} \right],
\end{equation}
where the different strength of two-particle creation and annihilation processes makes the Hamiltonian nonreciprocal. Collectively, these results establish how suitable choices of deformations $F_\alpha(k)$ generate a broad class of quadratic non-Hermitian Hamiltonians in real space, encompassing short-range, long-range, and nonreciprocal couplings with tunable EP structure.

\textit{Lindblad embedding:---}
We now discuss an experimentally feasible route involving Lindblad embedding to realize the aforementioned non-Hermitian Hamiltonians as effective descriptions emerging from open quantum 
dynamics. The starting point is the fact that the density matrix of an open quantum system under Markovian approximation evolves according to the Gorini--Kossakowski--Sudarshan--Lindblad 
(GKSL) equation~\cite{Breuer2007}, which is characterized by a set of jump operators 
$\{\hat L_\mu\}$ satisfying the complete positivity requirement 
$\sum_\mu \hat L_\mu^\dagger \hat L_\mu \geq 0$. In the quantum trajectory 
formulation of the GKSL dynamics, the evolution conditioned on the absence of quantum jumps (the so-called ``no-click'' limit) is governed by an effective non-Hermitian Hamiltonian, $\hat H^{\text{nH}}= \hat{H}+\delta\hat{H};~\delta\hat{H}=-i\sum_\mu \hat L_\mu^\dagger \hat L_\mu$. This provides a route to engineer suitable jump operators 
$\{\hat L_\mu\}$ such that the no-click limit of the corresponding GKSL 
evolution generates the desired non-Hermitian deformation \footnote{Indeed, this recovers the desired non-Hermitian Hamiltonian up to a constant shift, which has no effect on the EP physics.}.

To illustrate the embedding, we construct the $\{\hat L_\mu\}$ explicitly for a representative $\delta \hat{H}$ and then discuss the generality of the construction. Consider $\delta \hat H_y = (\gamma_y/2) \sum_{j=1}^L \left[\hat c_j^\dagger \hat c_{j+1}^\dagger + \hat c_j \hat c_{j+1} \right] $, where the $j$-th site is coupled to the $(j+1)$-th site. Since $\delta \hat H_y$ is quadratic in fermionic operators, we consider a general linear combination of creation and annihilation operators on the coupled sites, $\hat{L}_j = \alpha \hat{c}_j + \beta \hat{c}_{j+1}+\gamma \hat{c}^\dagger_{j}+\delta \hat{c}^\dagger_{j+1}$. Evaluating $-i\sum_{j=1}^L\hat{L}^\dagger_j\hat{L}_j $ and comparing with the desired deformation $\delta\hat{H}_y$ yield $\beta^{*}\delta = \alpha^{*}\gamma = 0$. Choosing $\gamma=\beta=0$, the jump operator becomes
\begin{align}
    \hat{L}^\dagger_j\hat{L}_j
    &= |\delta|^2 \mathbb{I}
    + \left(|\alpha|^2 \hat{c}^\dagger_j\hat{c}_j
    - |\delta|^2 \hat{c}^\dagger_{j+1}\hat{c}_{j+1}\right)
    \nonumber\\
    &\quad
    + \left(\alpha^{*}\delta\, \hat{c}^\dagger_j\hat{c}^\dagger_{j+1}
    - \delta^{*}\alpha\, \hat{c}_{j}\hat{c}_{j+1}\right),
\end{align}
which has the form of the desired $\delta \hat H_y$ provided $|\alpha|^2=|\delta|^2$ and $\alpha^{*}\delta=i\gamma_y/2$. A possible choice is $\alpha=\sqrt{\gamma_y/2}$ and $\delta=i\sqrt{\gamma_y/2}$, which gives
$\hat{L}_j=\sqrt{\gamma_y/2}\left(\hat{c}_j+i\hat{c}_{j+1}^{\dagger}\right)$. In general, if the non-Hermitian deformation $\delta\hat H$ couples any two sites $j$ and $m$ at arbitrary separation, one associates for the pair $(j,m)$ a jump operator  of the form
$\hat L_{j,m} =
    \alpha \hat c_j
    +\beta \hat c_m
    +\gamma \hat c_j^\dagger +\delta \hat c_m^\dagger ,
$
where the coefficients $\alpha,\beta,\gamma,\delta$ are determined by matching the terms generated from $-i\sum_{j,m}\hat L_{j,m}^\dagger\hat L_{j,m}$ with the desired $\delta\hat H$. This construction provides a simple recipe for realizing arbitrary non-Hermitian Hamiltonians within Lindblad dynamics.

\textit{Conclusion:--}
We established momentum space deformation as a general design principle for programmable non-Hermitian many-body systems. We identified universal criteria for the emergence of EP's, proved that a single momentum-space EP induces an exponential proliferation of pairwise many-body eigenvector coalescences, and showed that $\mathcal{PT}$ symmetry enables controlled purification of arbitrary mixed quantum states, with distinct purification regimes and a fundamental odd-even dichotomy in the thermodynamic limit. The framework extends beyond $2\times2$ momentum-space Hamiltonians to arbitrary discrete translational symmetries, where $D\times D$ BdG blocks support higher-order EP's, e.g., fourth-order EP's in the staggered-field TFIC ($D=4$). Our proposed Lindblad embedding provides a route to experimental realization; extension to Liouvillian EP's~\cite{Sun2024} via third quantization~\cite{Prosen2008} is a promising direction. We anticipate our results to stimulate studies of correlated non-Hermitian phases~\cite{Ashida2020}, dissipative quantum control~\cite{Harrington2022}, information processing and metrology~\cite{Verstraete2009,Shi2025} based on engineered exceptional-point physics.

\textit{Acknowledgement:--}
This work was supported by the Department of Atomic Energy, Government of India, under Project Identification Number RTI-4012. The computations were carried out on the computing clusters at the Department of Theoretical Physics, TIFR, Mumbai. We also thank Ajay Salve and Kapil Ghadiali for their computational support.

\bibliographystyle{apsrev4-2}
\bibliography{main.bib}



\clearpage
\clearpage
\appendix
\section{End Matter}
\section{Appendix A: Evolution under an $(n\times n)$ non-Hermitian matrix with EP of order $n$ can purify any arbitrary mixed state}

\underline{\textit{Claim:--}} If an $n \times n$ non-Hermitian matrix $\hat{H}$ hosts an EP of order $n$ (whereby the eigenvalue $\lambda$ becomes $n$-fold degenerate with only a single eigenvector), then any arbitrary mixed state under its evolution gets completely purified.  \\

\underline{\textit{Proof:---}} For $\hat H$ to host an EP of order $n$, it has to be related to the Jordan canonical form via a similarity transformation $S$ as 
\begin{align}
    \hat H = S ( \lambda \mathbb{I}_n + J_n)S^{-1},
\end{align}
where $[J_n]_{ij} =\delta_{i+1,j}$ is the highest order Jordan canonical form such that $(J_n)^n =0$.

Now, the time-evolution operator for $\hat H$ is 
\begin{align}
    U (t) &= e^{-i\hat H t } = S e^{-it (\lambda \mathbb{I}_n + J_n ) } S^{-1}  = e^{-it \lambda}  S e^{-it  J_n  } S^{-1} \nonumber\\
    & = e^{-it \lambda} S \biggr( \sum_{m=0}^{n-1} \frac{(-it)^m}{ m! }  (J_n)^m \biggr)   S^{-1}
\end{align}
which will lead to the evolved density matrix as 
\begin{align}
\hat \rho(t) & = \frac{U(t) \hat \rho_0U^\dagger(t)} {Tr\biggr[ U(t) \hat \rho_0U^\dagger(t) \biggr]} = S \biggr( \sum_{m}^{n-1} \frac{(-it)^m}{ m! }  (J_n)^m \biggr)   S^{-1} \nonumber \\
&  \times \frac{ \hat\rho_0 S \biggr( \sum_{m=0}^{n-1} \frac{(it)^m}{ m! }  (J_n^\dagger)^m \biggr)   S^{-1}}{Tr\biggr[ U(t) \hat \rho_0U^\dagger(t) \biggr]}.
\end{align}
Clearly, it is clear from the expression above that in the limit $t\gg1$, the dominating term for the density matrix comes directly from the term corresponding to $t^{n-1}$. Now, defining $\tilde J \equiv S J_n S^{-1}$, we write 
\begin{align}
    \hat \rho(t)  &= \frac{(it)^{2n-2} \tilde{J}^{n-1} \hat \rho_0 (\tilde{J}^\dagger)^{n-1} + \mathcal{O}(t^{2n-3}) }{ Tr\biggr[ (it)^{2n-2} \tilde{J}^{n-1} \hat \rho_0 (\tilde{J}^\dagger)^{n-1} + \mathcal{O}(t^{2n-3}) \biggr] } \\
    \implies &  \hat \rho(t \to \infty ) =\frac{ \tilde{J}^{n-1} \hat \rho_0 (\tilde{J}^{n-1})^\dagger }{ Tr\biggr[ \tilde{J}^{n-1} \hat \rho_0 (\tilde{J}^{n-1})^\dagger  \biggr] }.   
\end{align}
Now, by construction, $J_n$ is a matrix of rank-$(n-1)$, and rank is invariant under similarity transformation; therefore, $\tilde J$ is of rank-$(n-1)$. This leads to $\tilde J^{n-1}$ to be of rank-1. We know that any rank-$1$ matrix can be written as a dyad product, leading to $ \tilde J^{n-1} = |u \rangle \langle v| $. Putting this back in the asymptotic expression of the density matrix, we obtain 
\begin{align}
    \hat \rho(t \to \infty ) = \frac{ \langle v | \hat \rho_0 | v \rangle | u\rangle \langle u |  }{\langle v | \hat \rho_0 | v \rangle} = | u\rangle \langle u|.
\end{align}
Therefore, clearly $\hat \rho(t\to \infty)$ is a pure state, and as a corollary, EP always purifies arbitrary density matrices.

\section{Appendix B:~Purification Dynamics}

Let us consider the non-Hermitian Hamiltonian
\begin{equation}
    \hat{H} =  \sum_{k>0} \Psi_k^\dagger  \mathcal{H}_k^{\mathrm{nH}} \Psi_k 
\end{equation}
where we have $\mathcal{H}_k^{\mathrm{nH}}= (a_0-ib_0)\mathbb{I}+(\vec{a}-i\vec{b})\cdot \vec{\sigma}$. Time evolution operator under this Hamiltonian has the form
\begin{align}
    &e^{-i\hat Ht} = \bigotimes_{k>0} e^{-i \mathcal{H}_k^{\mathrm{nH}}t} \nonumber\\
    &=  \bigotimes_{k>0} e^{-i(a_0-ib_0)t}\bigg[\cos{(\omega_k t)}\mathbb{I}-i\frac{\sin{(\omega_k t)}}{\omega_k }(\vec{a}-i\vec{b})\cdot \vec{\sigma}\bigg].
\end{align}
where $\omega_k = \sqrt{a^2-b^2-i2\vec{a}\cdot\vec{b}}$. Let the dynamics be initiated at the state
\begin{equation}
    \rho(0) =\bigotimes_{k>0} \rho_k(0) =  \bigotimes_{k>0} \frac{1}{2}\Big(\mathbb{I}+\vec{p}_k\cdot\vec{\sigma}\Big),
\end{equation}
then the un-normalized density matrix at a time $t$ is given by
\begin{equation}
     \rho(t) \equiv  e^{-i\hat Ht}  \rho(0) e^{i\hat H^\dagger t}= \bigotimes_{k>0} \rho_k(t) =  \bigotimes_{k>0} e^{-i \mathcal{H}_k^{\mathrm{nH}}t}\rho_k(0) e^{i \mathcal{H}_k^{\mathrm{nH}^\dagger}t}.
\end{equation}
Straightforward computation shows $\rho_k(t)$ has the form $\rho_k(t) = 2^{-1}\Big[p_0(t)\mathbb{I}+\vec{p}_k(t)\cdot\vec{\sigma}\Big]$, where we have
\begin{align}
    p_0(t) &= \big|\mathcal{C}\big|^2+\big|\mathcal{S}\big|^2\bigg[a^2+b^2+2\big(\vec{a}\times\vec{b}\big)\cdot \vec{p}_k\bigg]+2 \Im\big[\mathcal{C}^{*}\mathcal{S}\big]\vec{a}\cdot \vec{p}_k \nonumber\\
    &-2 \Re\big[\mathcal{C}^{*}\mathcal{S}\big]\vec{b}\cdot \vec{p}_k, \label{eq: p_0(t)}
\end{align}
and
\begin{align}
    \vec{ p}_k(t) &= \big|\mathcal{C}\big|^2\vec{p}_k+\big|\mathcal{S}\big|^2\bigg\{2\big(\vec{a}\cdot\vec{p}_k\big)\vec{a}+2\big(\vec{b}\cdot\vec{p}_k\big)\vec{b} -\big(a^2+b^2\big)\vec{p}_k \nonumber\\
    &-2\big(\vec{a}\times\vec{b}\big)\bigg\}+2 \Im\big[\mathcal{C}^{*}\mathcal{S}\big]\vec{a}-2 \Re\big[\mathcal{C}^{*}\mathcal{S}\big]\vec{b}+2 \Re\big[\mathcal{C}^{*}\mathcal{S}\big]\vec{a}\times \vec{p}_k\nonumber\\
    &+2 \Im\big[\mathcal{C}^{*}\mathcal{S}\big]\vec{b}\times \vec{p}_k, \label{eq: p vec (t)}
\end{align}
where we have $\mathcal{C}  \equiv \cos{(\omega_k t)},~\mathcal{S}\equiv \sin{(\omega_k t)}/\omega_k ,~\vec{h} \equiv \vec{a}-i\vec{b}$. 

In case of a $\mathcal{PT}$-symmetric Hamiltonian, we have $\vec{a}\cdot \vec{b}=0$. This immediately gives $\omega_k = \sqrt{a^2-b^2}$, resulting in $\omega_k$ being either purely real or purely imaginary. In the $\mathcal{PT}$-symmetric phase $(a>b)$, $\omega_k$ is real, so that the purity $P_k(t) = \vec{p}_k(t) \cdot \vec{p}_k(t)/|p_0(t)|^2$ keeps on oscillating with time. In the $\mathcal{PT}$-symmetry broken phase $(b>a)$, $\omega_k$ is purely imaginary: $\omega_k = i\sqrt{b^2-a^2} = i\Omega_k$, where $\Omega_k$ is purely real. Hence, at long times, we have $\mathcal{C} = \mathcal{S} \approx e^{\Omega_kt}/2$. This results in the steady state normalized density matrix
\begin{equation}
    \rho_k(t\to \infty) = \frac{1}{2b}\begin{pmatrix}
        (b-a) & i\sqrt{b^2-a^2}\\
        -i\sqrt{b^2-a^2} & (b+a)
    \end{pmatrix},
\end{equation}
which is independent of the initial density matrix, and which has purity equal to unity. Hence, dynamics in the $\mathcal{PT}$-symmetry broken phase purifies arbitrary mixed state. For maximally mixed state, which has $\vec{p}_k=\vec{0}$, combining Eqs.~\eqref{eq: p_0(t)} and~\eqref{eq: p vec (t)} and computing purity, we get back Eq.~\eqref{eq: purity} in the main text.

\section{Appendix C: Results corresponding to Fig.~\ref{fig: fig 1}, but for odd $L$}

\begin{figure}[h!]
    \centering
    \includegraphics[width=0.8\linewidth]{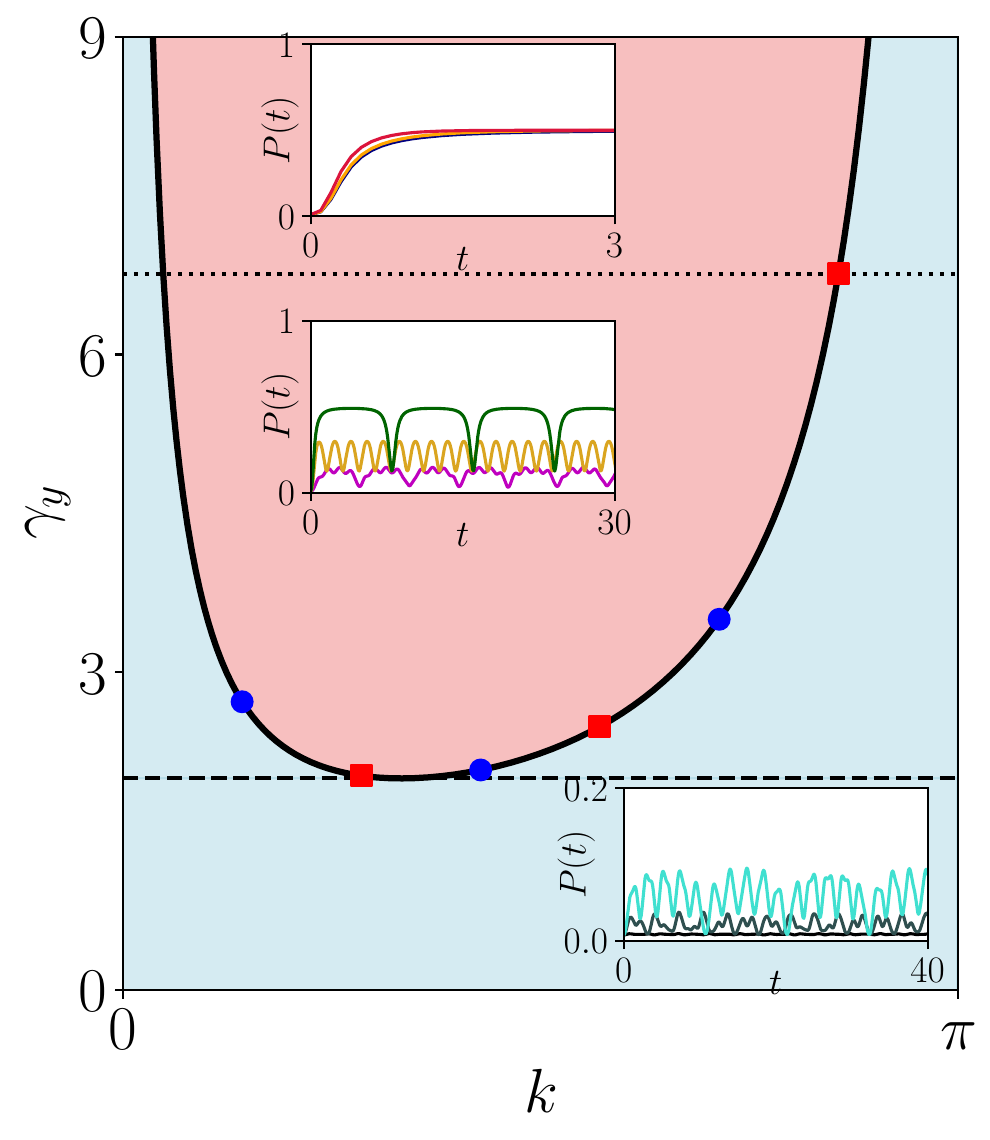}
    \caption{For the non-Hermitian Hamiltonian $H=H^{(1)}-iH^{(2)}$, with $H^{(1)}$ in Eq.~\eqref{eq: TFIC} and the $y$-deformation
$H^{(2)}=\gamma_y\sin k\,\sigma^y$, every allowed momentum block hosts an EP at $k_c=k\in\mathrm{RBZ}$. For $h=0.5$, $J=1$, the solid black curve shows $\gamma_y^{(c)}(k)$ in the thermodynamic limit, with the dashed horizontal line indicating $\gamma_\mathrm{c}^{(1)}$. For a finite system with $L=7$, the red (blue) circles denote the odd-parity (even-parity) momentum blocks. Although the upper threshold $\gamma_\mathrm{c}^{(2)}\equiv\max_{k\in\mathrm{RBZ}}\gamma_y^{(c)}(k)$ diverges in the thermodynamic limit, its finite-size value is indicated by the dotted horizontal line. Going from top to bottom, the three subpanels show the time evolution of the many-body purity, starting from the maximally mixed state, in the regimes (i) $\gamma_y (=0.5, 1.5,1.99)<\gamma_\mathrm{c}^{(1)}$, (ii) $\gamma_\mathrm{c}^{(1)}<\gamma_y(=2.4, 5.0,6.7)<\gamma_\mathrm{c}^{(2)}$, and (iii) $\gamma_y(=6.77,7.0,8.0)>\gamma_\mathrm{c}^{(2)}$.}
    \label{fig:Odd-L}
\end{figure}

\section{Appendix D: Purification under a Long range Hamiltonian}

\begin{figure}[h!]
    \centering
    \includegraphics[width=1.0\linewidth]{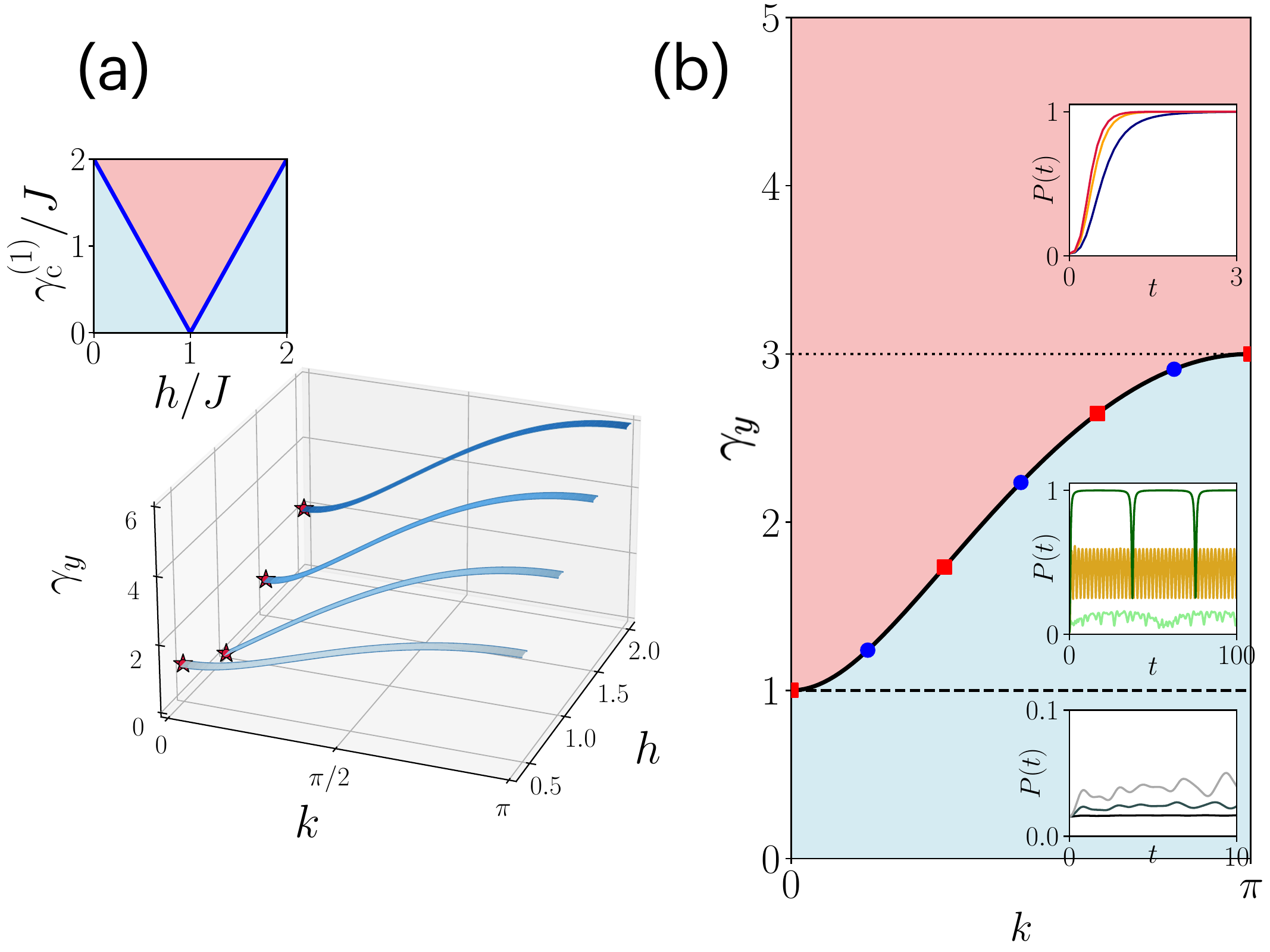}
    \caption{For the non-Hermitian Hamiltonian $H=H^{(1)}-iH^{(2)}$, with $H^{(1)}$ in Eq.~\eqref{eq: TFIC} and the $y$-deformation
$H^{(2)}=\gamma_y\sigma^y$, every allowed momentum block hosts an EP at $k_c=k\in\mathrm{RBZ}$. (a) Critical deformation strength $\gamma_y^{(c)}$ in Eq.~\eqref{eq: gamma c} as a function of $(h,k,\gamma_y)$. The stars mark the minimum threshold $\gamma_\mathrm{c}^{(1)}\equiv\min_{k\in\mathrm{RBZ}}\gamma_y^{(c)}(k)$, while the inset shows $\gamma_\mathrm{c}^{(1)}/J$ versus $h/J$. (b) For $h=0.5$, $J=1$, the solid black curve shows $\gamma_y^{(c)}(k)$ in the thermodynamic limit, with the dashed (respectively, dotted) horizontal line indicating $\gamma_\mathrm{c}^{(1)}$ (respectively, $\gamma_\mathrm{c}^{(2)}$). For a finite system with $L=6$, the red (blue) circles denote the odd-parity (even-parity) momentum blocks. Going from top to bottom, the three subpanels show the time evolution of the many-body purity, starting from the maximally mixed state, in the regimes (i) $\gamma_y (=0.2, 0.6, 0.9)<\gamma_\mathrm{c}^{(1)}$, (ii) $\gamma_\mathrm{c}^{(1)}<\gamma_y(=1.4,2.4,2.877)<\gamma_\mathrm{c}^{(2)}$, and (iii) $\gamma_y(=3.1,4.0,4.5)>\gamma_\mathrm{c}^{(2)}$.}
    \label{fig:placeholder}
\end{figure}

\clearpage

\onecolumngrid

\newpage

\begin{center}

{\large \bf Supplemental Material: ``Constructing Non-Hermitian Theories with Tunable Exceptional Points and Controlled State Purification" 
\\--------}\\

\vspace{0.6cm}

Soumya Kanti Pal, Rupak Majumder, Shamik Gupta\\

$^1${\it Department of Theoretical Physics, Tata Institute of Fundamental Research, Homi Bhabha Road, Mumbai 400005, India}

\end{center}

\vspace{0.6cm}


\setcounter{section}{0}
\renewcommand\thesection{\Roman{section}}
\renewcommand\thesubsection{\thesection.\Alph{subsection}}
\renewcommand{\thefigure}{S\arabic{figure}}
\renewcommand{\theequation}{S\arabic{equation}}

\section{Derivation of Eq.~(3) of the main text}

As discussed in the main text, any $(2\times 2)$ non-Hermitian matrix  $  H_k^{\mathrm{nH}} \neq (H_k^{\mathrm{nH}})^\dagger$ admits the decomposition 
\begin{align}
    \mathcal{H}_k^{\text{nH}} = H^{(1)} - i H^{(2)},
\end{align}
where $H^{(1)},~ H^{(2)}$ are two Hermitian matrices with the following parametrization
\begin{align}
    H^{(1)} = a_0 \mathbb{I} + \vec{a}\cdot \vec{\sigma},~~ H^{(2)} = b_0 \mathbb{I} + \vec{b}\cdot \vec{\sigma},
\end{align}
where $a_0, b_0 \in \mathbb{R}$, and $\vec{a},\vec{b} \in \mathbb{R}^3$. An EP of $\mathcal{H}_k^{\mathrm{nH}}$ occurs when it possesses a doubly degenerate eigenvalue $\lambda$ and is simultaneously defective, i.e., it admits only one linearly independent eigenvector. This tantamounts to the matrix
$B \equiv \mathcal{H}_k^{\mathrm{nH}} - \lambda \mathbb{I}_2$ becoming \textit{similar} to the nilpotent Jordan block,
\begin{align}
    B = S J_2(0) S^{-1}, \qquad
    J_2(0) =
    \begin{pmatrix}
        0 & 1 \\
        0 & 0
    \end{pmatrix},
\end{align}
where $S$ is an invertible similarity transformation, and $J_2(0)$ corresponds to the fundamental Jordan block with eigenvalue $0$. Consequently,
\begin{align}
    B^2 = 0, \qquad B \neq 0,
\end{align}
which provides the algebraic characterization of a second-order exceptional point. 

Now, from the Pauli basis decomposition of $\mathcal{H}_k^{\mathrm{nH}}$, the two eigenvalues of the matrix are $\lambda_{\pm} = (a_0 - ib_0) \pm \sqrt{|a|^2 - |b|^2 - 2i \vec a \cdot \vec b} $. From the degeneracy condition, we get,  $|a|^2 - |b|^2 - 2i \vec a \cdot \vec b = 0 $, which can be further reduced by separating the real and the imaginary part as 
\begin{align}
    (i)~ \vec a \cdot \vec b = 0, ~(ii) ~ |\vec a | = | \vec b | . 
\end{align}
These conditions automatically guaranty that $B^2=0$, since at EP, $B^2 =\left( (\vec a - i \vec b ) \cdot \sigma \right) \left( (\vec a - i \vec b ) \cdot \sigma \right) = | \vec a |^2 - | \vec b|^2 - 2i \vec a \cdot \vec b$. However, the solution $\vec a = \vec b$, is not allowed since it violates $B \neq 0$ condition. Therefore, these conditions together constitute the EP conditions in Eq.~(3) of the main text.

\section{$\mathcal{PT}$-symmetric non-Hermitian Hamiltonian}

In a spin chain, parity symmetry usually means reflection of the lattice about its center. For a chain of $L$ spins with periodic boundary condition, the parity operator $\hat{\mathcal{P}}$ is defined as
\begin{equation}
    \hat{\mathcal{P}}\sigma_j^\alpha\hat{\mathcal{P}}^{-1} = \sigma_{L+1-j}^\alpha,~\forall~j
\end{equation}
where $\alpha = x,y,z$. We now find the corresponding relation in the spinless fermion picture, related to the Pauling spin matrices by the Jordan-Wigner transformation, defined as
\begin{align}
    c_{j} &\equiv \frac{1}{2}\left[\prod_{k=1}^{j-1}\sigma^z_{k}\right]\Big(\sigma^x_j+i\sigma^y_j\Big) ,\label{eq: c_j}\\
     c^\dagger_{j} &\equiv \frac{1}{2}\left[\prod_{k=1}^{j-1}\sigma^z_{k}\right]\Big(\sigma^x_j-i\sigma^y_j\Big). \label{eq: c dagger_j}
\end{align}
We can invert these equations and obtain the following equations
\begin{align}
    \sigma^x_j &= \left[\prod_{k=1}^{j-1}\sigma^z_{k}\right]\Big( \hat{c}^\dagger_{j} + \hat{c}_{j} \Big) ,\\
     \sigma^y_j &=i \left[\prod_{k=1}^{j-1}\sigma^z_{k}\right]\Big( \hat{c}^\dagger_{j} - \hat{c}_{j} \Big) ,\\
     \sigma^z_j &= 1 - 2\hat{c}^\dagger_j\hat{c}_j.
\end{align}
Applying Parity operator in both sides of Eq.~\eqref{eq: c_j}, we obtain that
\begin{align}
    \hat{\mathcal{P}}\hat{c}_{j}\hat{\mathcal{P}}^{-1} &\equiv \frac{1}{2}\left[\prod_{k=1}^{j-1}\sigma^z_{L+1-k}\right]\Big(\sigma^x_{L+1-j}+i\sigma^y_{L+1-j}\Big) = \left[\prod_{k=1}^{j-1}\sigma^z_{L+1-k}\right]\left[\prod_{m=1}^{L-j}\sigma^z_m\right]\hat{c}_{L+1-j}\nonumber\\
    & = \left[\prod_{m=L-j+2}^{L}\sigma^z_m\right]\left[\prod_{m=1}^{L-j}\sigma^z_m\right]\hat{c}_{L+1-j} = \left[\prod_{m=1}^{L}\sigma^z_m\right] \sigma^z_{L+1-j}\hat{c}_{L+1-j} = (-1)^{\hat{N}_f} \hat{c}_{L+1-j}, 
\end{align}
where, along with the definition $\hat{N}_f \equiv \sum_{j=1}^L \hat{c}^\dagger_j\hat{c}_j$, we have used the fact that $\sigma^z_{L+1-j}\hat{c}_{L+1-j} = (1 - 2\hat{c}^\dagger_{L+1-j}\hat{c}_{L+1-j})\hat{c}_{L+1-j} = \hat{c}_{L+1-j}$. Similarly, applying Parity operator in both sides of Eq.~\eqref{eq: c dagger_j} and following similar procedure, we obtain that
\begin{align}
    \hat{\mathcal{P}}\hat{c}^\dagger_{j}\hat{\mathcal{P}}^{-1} &\equiv \frac{1}{2}\left[\prod_{k=1}^{j-1}\sigma^z_{L+1-k}\right]\Big(\sigma^x_{L+1-j}-i\sigma^y_{L+1-j}\Big) = \left[\prod_{k=1}^{j-1}\sigma^z_{L+1-k}\right]\left[\prod_{m=1}^{L-j}\sigma^z_m\right]\hat{c}^\dagger_{L+1-j}\nonumber\\
    & = \left[\prod_{m=L-j+2}^{L}\sigma^z_m\right]\left[\prod_{m=1}^{L-j}\sigma^z_m\right]\hat{c}^\dagger_{L+1-j} = \left[\prod_{m=1}^{L}\sigma^z_m\right] \sigma^z_{L+1-j}\hat{c}^\dagger_{L+1-j} = -(-1)^{\hat{N}_f} \hat{c}^\dagger_{L+1-j}. 
\end{align}

We now compute the action of the Parity operator on the spinless fermionic operators in the momentum space. Using the definition, we obtain
\begin{align}
    \hat{\mathcal{P}}\hat{c}_k\hat{\mathcal{P}}^{-1} &= \frac{1}{\sqrt{L}}\sum_{j = 1}^L e^{-ikj}   \hat{\mathcal{P}}\hat{c}_j   \hat{\mathcal{P}}^{-1} = (-1)^{\hat{N}_f} \frac{1}{\sqrt{L}}\sum_{j = 1}^L e^{-ikj}   \hat{c}_{L+1-j} = (-1)^{\hat{N}_f} e^{-ik(L+1)} \frac{1}{\sqrt{L}}\sum_{j = 1}^L e^{ik(L+1-j)}   \hat{c}_{L+1-j} \nonumber \\
    &=  (-1)^{\hat{N}_f} e^{-ik(L+1)}\hat{c}_{-k}.
\end{align}
Doing a similar computation, we obtain
\begin{align}
    \hat{\mathcal{P}}\hat{c}^\dagger_k\hat{\mathcal{P}}^{-1} &= \frac{1}{\sqrt{L}}\sum_{j = 1}^L e^{ikj}   \hat{\mathcal{P}}\hat{c}^\dagger_j   \hat{\mathcal{P}}^{-1} = -(-1)^{\hat{N}_f}  \frac{1}{\sqrt{L}}\sum_{j = 1}^L e^{ikj}   \hat{c}^\dagger_{L+1-j} = -(-1)^{\hat{N}_f} e^{ik(L+1)} \frac{1}{\sqrt{L}}\sum_{j = 1}^L e^{-ik(L+1-j)}   \hat{c}^\dagger_{L+1-j} \nonumber \\
    &=  -(-1)^{\hat{N}_f} e^{ik(L+1)}\hat{c}^\dagger_{-k}.
\end{align}

We now focus on computing the action of the time reversal operator $\hat{\mathcal{T}}$, which is simply the complex conjugation operator with the defining property $\hat{\mathcal{T}}i\hat{\mathcal{T}}^{-1} = -i$. Hence, the action of $\hat{\mathcal{T}}$ on the real space Pauling spin operators are
\begin{equation}
    \hat{\mathcal{T}}\hat{\sigma}^x\hat{\mathcal{T}}^{-1} = \hat{\sigma}^x,~~\hat{\mathcal{T}}\hat{\sigma}^y\hat{\mathcal{T}}^{-1} = -\hat{\sigma}^y,~~\hat{\mathcal{T}}\hat{\sigma}^z\hat{\mathcal{T}}^{-1} = \hat{\sigma}^z.
\end{equation}
Applying the $\hat{\mathcal{T}}$ in both sides of Eq.~\eqref{eq: c_j}, we obtain that
\begin{align}
    \hat{\mathcal{T}}\hat{c}_{j}\hat{\mathcal{T}}^{-1} &= \frac{1}{2}\left[\prod_{k=1}^{j-1}\sigma^z_{k}\right]\Big(\sigma^x_{j}+i\sigma^y_{j}\Big) = \hat{c}_{j}.
\end{align}
Similarly, applying the $\hat{\mathcal{T}}$ in both sides of Eq.~\eqref{eq: c dagger_j}, we obtain
\begin{align}
    \hat{\mathcal{T}}\hat{c}^\dagger_{j}\hat{\mathcal{T}}^{-1} &= \frac{1}{2}\left[\prod_{k=1}^{j-1}\sigma^z_{k}\right]\Big(\sigma^x_{j}-i\sigma^y_{j}\Big) = \hat{c}^\dagger_{j}.
\end{align}

We now compute the action of $\hat{\mathcal{T}}$ on the spinless fermionic operators in the momentum space. Using the definition, we obtain
\begin{align}
    \hat{\mathcal{T}}\hat{c}_k\hat{\mathcal{T}}^{-1} &= \hat{\mathcal{T}}\Bigg[\frac{1}{\sqrt{L}}\sum_{j = 1}^L e^{-ikj}   \hat{c}_j   \Bigg]\hat{\mathcal{T}}^{-1} = \frac{1}{\sqrt{L}}\sum_{j = 1}^L e^{ikj}   \hat{c}_j = \hat{c}_{-k},
\end{align}
and similarly
\begin{align}
    \hat{\mathcal{T}}\hat{c}^\dagger_k\hat{\mathcal{T}}^{-1} &= \hat{\mathcal{T}}\Bigg[\frac{1}{\sqrt{L}}\sum_{j = 1}^L e^{ikj}   \hat{c}^\dagger_j   \Bigg]\hat{\mathcal{T}}^{-1} = \frac{1}{\sqrt{L}}\sum_{j = 1}^L e^{-ikj}   \hat{c}^\dagger_j = \hat{c}^\dagger_{-k}.
\end{align}

We now have all the components to find out the conditions for a $\mathcal{PT}$-symmetric non-Hermitian Hamiltonian. In our convention $\Psi_k^\dagger = \begin{pmatrix}
    c_k^\dagger & i c_{-k}
\end{pmatrix}$. If we denote
\begin{align}
    \mathcal{H}^\mathrm{nH}_k = \begin{pmatrix}
    A_\mathrm{nH}(k) & B_\mathrm{nH}(k)\\
    C_\mathrm{nH}(k) & D_\mathrm{nH}(k)
\end{pmatrix},
\end{align}
then $k$ momentum block Hamiltonian is 
\begin{equation}
    \Psi_k^\dagger \mathcal{H}^\mathrm{nH}_k \Psi_k = A_\mathrm{nH} c_k^\dagger c_k-iB_\mathrm{nH} c_k^\dagger c_{-k}^\dagger  + i C_\mathrm{nH} c_{-k} c_{k} + D_\mathrm{nH} c_{-k}c_{-k}^\dagger.
\end{equation}
Application of the $\mathcal{PT}$-symmetry operator gives
\begin{equation}
    \mathcal{PT}\Big[\Psi_k^\dagger \mathcal{H}^\mathrm{nH}_k \Psi_k\Big]\mathcal{T}^{-1}\mathcal{P}^{-1} = A^*_\mathrm{nH} c_k^\dagger c_k-iB^*_\mathrm{nH} c_k^\dagger c_{-k}^\dagger  + i C^*_\mathrm{nH} c_{-k} c_{k} + D^*_\mathrm{nH} c_{-k}c_{-k}^\dagger.
\end{equation}
If the Hamiltonian is $\mathcal{PT}$ symmetric, we have
\begin{equation}
    \mathcal{PT}\Big[\Psi_k^\dagger \mathcal{H}^\mathrm{nH}_k \Psi_k\Big]\mathcal{T}^{-1}\mathcal{P}^{-1}  = \Psi_k^\dagger \mathcal{H}^\mathrm{nH}_k \Psi_k.
\end{equation}
Comparing them, we obtain that, if a Hamiltonian is $\mathcal{PT}$, it must satisfy
\begin{equation}
    \mathcal{H}^\mathrm{nH}_k = \Big[\mathcal{H}^\mathrm{nH}_k\Big]^*.
\end{equation}
Now for the $\mathcal{H}^\mathrm{nH}_k = (\vec a-i\vec b ) \cdot \hat {\vec \sigma}$ considered in the main text, the above condition gives us 
\begin{align}
    \mathcal{H}^\mathrm{nH}_k = a_0 \mathbb{I} + a_x \sigma^x -ib_y \sigma^y + a_z \sigma^z.  
\end{align}

\section{Recap of some relevant properties of TFIC} \label{Sec:Derivation of Eq.~(7)}

We begin with the one-dimensional periodic transverse-field Ising model in the spin representation, described by
\begin{align}
\hat{H}
=
-J\sum_{j=1}^{L}\hat{\sigma}_{j}^{x}\hat{\sigma}_{j+1}^{x}
-h\sum_{j=1}^{L}\hat{\sigma}_{j}^{z},
\label{eq:Hamiltonian-TFIM}
\end{align}
subject to the periodic boundary condition $\hat{\sigma}_{N+1}^{x}=\hat{\sigma}_{1}^{x}$. To map the model onto spinless fermions, we employ the Jordan-Wigner transformation by defining 
\begin{align}
    \hat{c}_{j} \equiv \left[\prod_{k<j}\hat{\sigma}^z_{k}\right]\hat{\sigma}^+_j,\quad
     \hat{c}^\dagger_{j} \equiv \left[\prod_{k<j}\hat{\sigma}^z_{k}\right]\hat{\sigma}^-_j; \quad\hat{\sigma}^{\pm}_j\equiv \frac{\hat{\sigma}^x_j \pm i\hat{\sigma}^y_j}{2};
\end{align}
which follows the anticommutation relations $\{\hat{c}_i,\hat{c}^\dagger_j\}=\delta_{ij},~\{\hat{c}_i,\hat{c}_j\}=0,~\forall~ij$ with $(\hat{c}_j^\dagger)^2= (\hat{c}_j)^2=0$. Inverting the above transformations, the spin operators can be written entirely in terms of the Fermionic operators as
\begin{equation}
\hat{\sigma}_{j}^{z} = 1-2\hat{c}_{j}^{\dagger}\hat{c}_{j},\qquad
\hat{\sigma}_{j}^{+}=\hat{c}_{j}e^{i\pi\sum_{\ell<j}\hat{n}_{\ell}},\qquad
\hat{\sigma}_{j}^{-}=\hat{c}_{j}^{\dagger}e^{-i\pi\sum_{\ell<j}\hat{n}_{\ell}},
\end{equation}
where $\hat{n}_{j}=\hat{c}_{j}^{\dagger}\hat{c}_{j}$. Combining $\hat{\sigma}_{j}^{\pm}$ and noting that $e^{i\pi \hat{n}_j } = e^{-i\pi \hat{n}_j } = \hat\sigma_i^z$, we may write $\hat{\sigma}_i^x = (\hat{c}_i^\dagger + \hat{c}_i ) \prod_{j<i} \hat\sigma_j^z$, which immediately gives for the bulk indices ($i<L$) 
\begin{equation}
    \hat\sigma_i^x \hat\sigma_{i+1}^x  = (\hat{c}_i^\dagger - \hat{c}_i ) (\hat{c}_{i+1}^\dagger + \hat{c}_{i+1} ).
\end{equation}
The boundary term becomes
\begin{align}
\hat{\sigma}_{L}^{x}\hat{\sigma}_{1}^{x}
=
-\hat{\mathcal{P}}_f
(\hat{c}_{L}^{\dagger}-\hat{c}_{L})
(\hat{c}_{1}^{\dagger}+\hat{c}_{1}),
\end{align}
where $\hat{\mathcal{P}}_f= \prod_{j=1}^{N}\hat{\sigma}_{j}^{z}$ is the fermion-parity operator satisfying $[\hat{\mathcal{P}}_f,\hat{H}]=0$. The Hamiltonian therefore assumes the quadratic fermionic form
\begin{align}
\hat{H}
=
-J\left[
\sum_{j=1}^{L-1}
(\hat{c}_{j}^{\dagger}-\hat{c}_{j})
(\hat{c}_{j+1}^{\dagger}+\hat{c}_{j+1})
-
\hat{\mathcal{P}}_f
(\hat{c}_{L}^{\dagger}-\hat{c}_{L})
(\hat{c}_{1}^{\dagger}+\hat{c}_{1})
\right]
-h\sum_{j=1}^{L}
\left(1-2\hat{c}_{j}^{\dagger}\hat{c}_{j}\right). \label{eq:TFICin Fermion1}
\end{align}
Using the properties of the Fermionic operators, we may write $1-2\hat{c}_{j}^{\dagger}\hat{c}_{j} =\hat{c}_{j}\hat{c}^{\dagger}_{j} -\hat{c}_{j}^{\dagger}\hat{c}_{j}$. Defining $\hat{c}_{L+1} \equiv -\hat{\mathcal{P}}_f \hat{c}_1$, we may thus rewrite Eq.~\eqref{eq:TFICin Fermion1} as
\begin{equation}
    \hat{H} = -J \sum_{i=1}^{L}\Big[\hat{c}_i^\dagger \big(\hat{c}_{i+1}^\dagger+\hat{c}_{i+1}\big)+\mathrm{h.c.}\Big]+2h\sum_{i=1}^{L}\hat{c}_i^\dagger \hat{c}_i -Lh\label{eq:TFIC_SUP}.
\end{equation}
Equation~\eqref{eq:TFIC_SUP} is the Eq.~(7) of the main text upto a constant shift.

Since $\hat{\mathcal{P}}_f^2 = \mathbb{I}$ and $\hat{\mathcal{P}}_f^\dagger = \hat{\mathcal{P}}_f$, the eigenvalues can only be $\pm 1$.  We first focus on the eigenvalue $+1$, known as the even fermionic-parity sector. In this case we have $\hat{c}_{L+1} \equiv - \hat{c}_1$, which enables to write the Hamilotnian as
\begin{equation}
      \hat{\mathbb{P}}_{+1} \hat{H} \hat{\mathbb{P}}_{+1} = -J \sum_{i=1}^{L} (\hat{c}_i^\dagger - \hat{c}_i ) (\hat{c}_{i+1}^\dagger + \hat{c}_{i+1} )    - h \sum_{i=1}^L ( 1- 2 \hat{c}_i^\dagger \hat{c}_i   ), \label{eq: P1HP1}
\end{equation}
where $ \hat{\mathbb{P}}_{+1}$ is the projection operator in the even parity sector. We now go in the Fourier basis $\hat{c}_j = \sum_{k}e^{ijk} \hat{c}_k $. Performing Fourier transform for both $\hat c_{L+1}$ and $\hat c_1$ and putting it into $\hat{c}_{L+1} \equiv - \hat{c}_1$ we obtain that the momenta values $k$ must satisfy the following condition
\begin{equation}
    e^{iLk} = -1. \label{eq: k condition even parity}
\end{equation}
Equation~\eqref{eq: k condition even parity} restricts the allowed values of $k$ to
\begin{equation}
    k = \pm \frac{\pi}{L}, \pm \frac{3\pi}{L},\ldots , \pm \frac{(L-1)\pi}{L},
\end{equation}
for even $L$ and correspondingly
\begin{equation}
    k = \pm \frac{\pi}{L}, \pm \frac{3\pi}{L},\ldots , \pm \frac{(L-2)\pi}{L}, \pi,
\end{equation}
for odd $L$. Using these Fourier series, we may rewrite the following expressions as
\begin{equation}
    \sum_{j}\hat{c}_j^\dagger\hat{c}_{j+1}^\dagger = \sum_k e^{ik} \hat{c}_k^\dagger\hat{c}_{-k}^\dagger,\quad \sum_{j}\hat{c}_j^\dagger\hat{c}_{j+1} =\sum_k e^{ik} \hat{c}_k^\dagger\hat{c}_{k}. \label{eq: cjcj to ckck}
\end{equation}
Using Eq.~\eqref{eq: cjcj to ckck} into Eq.~\eqref{eq: P1HP1} we obtain for even $L$
\begin{equation}
     \hat{\mathbb{P}}_{+1} \hat{H} \hat{\mathbb{P}}_{+1} = \sum_{k>0} (2h - 2J \cos{k}) (\hat{c}_k^\dagger \hat{c}_k - \hat{c}_{-k} \hat{c}_{-k}^\dagger ) - \sum_{k>0} 2Ji \sin{k}~( \hat{c}_k^\dagger\hat{c}_{-k}^\dagger + \hat{c}_k\hat{c}_{-k}), \label{eq: P1HP1 in k}
\end{equation}
where $ \hat{\mathbb{P}}_{+1}$ is the projection operator in the even parity sector.  Defining Nambu-Gorkov spinors as $\Psi^\dagger_k = \begin{pmatrix}
        \hat{c}^\dagger_k  &
       i  \hat{c}_{-k}
    \end{pmatrix} $, we may rewrite Eq.~\eqref{eq: P1HP1 in k}
 as
 \begin{equation}
     \hat{\mathbb{P}}_{+1} \hat{H} \hat{\mathbb{P}}_{+1} = \sum_{k>0} \Psi_k^\dagger \hat{H}_k \Psi_k,
 \end{equation}
where we have
\begin{equation}
    \hat{H}_k   = 2 \begin{pmatrix}
        h - J \cos k & J \sin k \\
        J \sin{k} & -h + J \cos{k}
    \end{pmatrix} = 2 (h - J \cos{k} ) \hat\sigma^z + 2J \sin{k} \hat\sigma^x.
\end{equation}

We now focus on the sector corresponding to the $-1$ eigenvalue of $\hat{\mathcal{P}}_f$, known as the odd parity sector. In this case we have $\hat{c}_{L+1} \equiv  \hat{c}_1$ which puts the following condition on the momenta $k$:
\begin{equation}
    e^{iLk} = 1. \label{eq: k condition odd parity}
\end{equation}
Equation~\eqref{eq: k condition odd parity} restricts the allowed values of $k$ to
\begin{equation}
    k = 0, \pm \frac{2\pi}{L}, \pm \frac{4\pi}{L},\ldots , \pm \frac{(L-2)\pi}{L}, \pi,
\end{equation}
for even $L$ and correspondingly
\begin{equation}
    k =0, \pm \frac{2\pi}{L}, \pm \frac{4\pi}{L},\ldots , \pm \frac{(L-1)\pi}{L}, 
\end{equation}
for odd $L$. In this case, doing the similar computations as before, we obtain
\begin{equation}
    \hat{\mathbb{P}}_{-1} \hat{H} \hat{\mathbb{P}}_{-1} = \sum_{k>0} \Psi_k^\dagger \hat{H}_k \Psi_k + \frac{1}{2} \left(  \tilde\Psi_0^\dagger \hat{H}_0 \tilde\Psi_0 +  \tilde\Psi_{\pi}^\dagger \hat{H}_{\pi} \tilde\Psi_{\pi} \right) ,
\end{equation}
where $ \hat{\mathbb{P}}_{-1}$ is the projection operator in the even parity sector.
Thus we have explained the discussion following Eq.~(7) of the main text.



\section{Details on constructing various real-space non-Hermitian theories}
In this part, we discuss how deformations in individual directions in the momentum basis as $-i\gamma_\alpha F_{\alpha} (k)$ can be exploited to create a real-space non-Hermitian part featuring either short-range or long-range hopping, and subsequently mixing them to create a nonreciprocal theory.

\subsection{$\alpha=x$ -- deformation}
Corresponding to the momentum space deformation of the form $-i\gamma_x F_{x} (k)$, the total non-Hermitian deformation $\delta \hat H_x$ in the chosen Nambu-Gorkov representation takes the form 
\begin{align} \label{Eq:deformation-x}
    \delta\hat{H}_x & = -i \gamma_x \sum_{k>0} F_x(k) \tilde\Psi_k^\dagger \sigma^x \tilde\Psi_k = - \gamma_x \sum_{k>0} F_x(k) \hat{S}^x(k),
\end{align}
where $\hat{S}^x(k) \equiv \left( \hat{c}_k^\dagger \hat{c}_{-k}^\dagger - \hat{c}_{-k} \hat{c}_{k}   \right)$ is antisymmetric since $\hat{S}^x(k)= - \hat{S}^x(-k)$. This leads to two distinct situations for the choice of $F_x(k)$, coming from the fact that to get to real space, the sum over momenta in $\delta \hat H_x$ must be extended to the full Brillouin zone. In particular, (a) if  $F_x(k)$ is odd, then 
$\delta\hat{H}_x =- \frac{\gamma_x}{2}  \sum_{k\in BZ} F_x(k) \hat{S}^x(k)$, and (b) if $F(k) $ is even, then $\delta\hat{H}_x =- \frac{\gamma_x}{2}  \sum_{k\in BZ} \mathrm{sgn}(k) F_x(k) \hat{S}^x(k)$. Upon Fourier transformation, the former gives us a short-range hopping, while the latter creates a power-law hopping.

To demonstrate, we take $F_x(k)= \sin(mk) = -F_x(k)$, for which the $\delta \hat H_x$ in Eq.~\eqref{Eq:deformation-x} in real space becomes, 
\begin{align}
        \delta \hat{H}_x & =- \frac{\gamma_x}{4i} \sum_{jl} \left[ \frac{1}{L} \sum_k  e^{ik(j-l+m)}   \left( \hat{c}_j^\dagger\hat{c}_l^\dagger - \hat{c}_j\hat{c}_l\right)   - \frac{1}{L} \sum_k  e^{ik(j-l-m)}   \left\{ \hat{c}_j^\dagger\hat{c}_l^\dagger - \hat{c}_j\hat{c}_l\right\}  \right] \\
        & =- \frac{\gamma_x}{2i} \sum_{j=1}^L \left[ \hat{c}_j^\dagger\hat{c}_{j+m}^\dagger + \hat{c}_{j+m} \hat{c}_j\right],
    \end{align}  
with the condition in mind $ \hat c_{L+1} = \hat{\mathcal{P}}_f \hat c_1$. By tuning the harmonic order $m$, one can tune a controllable range of non-Hermitian hopping to its existing Hermitian counterpart.

Next, we demonstrate that the Fourier transform of $\mathrm{sgn}(k)$ for even $F_x(k)$ will lead to a hopping profile of the form $1/r$. For illustrative purposes, we take $F_x(k)=1$ and evaluate the non-Hermitian deformation in Eq.~\eqref{Eq:deformation-x} to be 
\begin{align}
    \delta \hat{H}_x & = - \frac{\gamma_x}{2} \sum_{jl} \mathcal{K}_{j,l}~ \left(\hat c_j^\dagger  \hat c_l^\dagger - \hat c_j \hat c_l \right),  
\end{align}
where we have 
\begin{align}
    \mathcal{K}_{jl} = \frac{1}{L}\sum_{k} e^{ik (j-l)}~\mathrm{sgn}(k) & = \frac{1}{L}\sum_{n=1}^{L/2 -1} \left( e^{2 i \pi n (j-l)/L} -  e^{-2 i \pi n (j-l)/L} \right) \\
    & = \begin{cases}
        0, ~ \text{if } (j-l) \text{ is even}\\
        \frac{2i}{L} \cot{\left( \frac{\pi (j-l)}{L}\right) }, ~ \text{if } (j-l) \text{ is odd}
    \end{cases}.
\end{align}
This finally gives us 
\begin{align}
    \delta \hat H_x = - \frac{i\gamma_x}{L } \sum_{i,j: (i-j)\in \text{Odd}} \cot{\left( \frac{\pi (j-l)}{L}\right) } \left(\hat c_j^\dagger  \hat c_l^\dagger - \hat c_j \hat c_l \right),
\end{align}
which in the large $L>>1$ limit takes the form 
\begin{align}
    \delta \hat H_x = - \frac{i\gamma_x}{\pi } \sum_{j,l: (j-l)\in \text{Odd}} \frac{1}{(j-l)} \left(\hat c_j^\dagger  \hat c_l^\dagger - \hat c_j \hat c_l \right) = - \frac{2i\gamma_x}{\pi } \sum_j \sum_{r>0, ~r\in odd} \frac{1}{r}\left[ \hat{c}_j^\dagger\hat{c}_{j+r}^\dagger + \hat{c}_j \hat{c}_{j+r} \right].
\end{align}
In general, a similar computation for $F_x(k) = \cos{(mk)}$ would reveal the same long-range structure in the non-Hermitian hopping structure. 

\subsection{$\alpha=y$- deformation}
Corresponding to the momentum space deformation of the form $-i\gamma_y F_{y} (k)$, the total non-Hermitian deformation $\delta \hat H_y$ in the chosen Nambu-Gorkov representation takes the form 
\begin{align} \label{Eq:deformation-x}
    \delta\hat{H}_y & = -i \gamma_y \sum_{k>0} F_y(k) \tilde\Psi_k^\dagger \sigma^y \tilde\Psi_k = -i \gamma_y \sum_{k>0} F_y(k) \hat{S}^y(k),
\end{align}
where $\hat{S}^y(k) \equiv \left( \hat{c}_k^\dagger \hat{c}_{-k}^\dagger + \hat{c}_{-k} \hat{c}_{k}   \right)$ is antisymmetric since $\hat{S}^y(k)= - \hat{S}^y(-k)$. Following the same argument as earlier, we get: (a) if  $F_y(k)$ is odd, then 
$\delta\hat{H}_y =- \frac{i\gamma_y}{2}  \sum_{k\in BZ} F_y(k) \hat{S}^y(k)$, and (b) if $F(k) $ is even, then $\delta\hat{H}_y =- \frac{i\gamma_y}{2}  \sum_{k\in BZ} \mathrm{sgn}(k) F_y(k) \hat{S}^y(k)$. As mentioned earlier, upon the inverse Fourier transformation, the former gives us a short-range hopping, while the latter creates a power-law hopping.

As earlier, we demonstrate by first  considering $F_y(k) = \sin{(mk)}$, which leads to the following short-range real-space Hamiltonian 
\begin{align}
    \delta \hat H_y = \frac{\gamma_y}{2} \sum_{j=1}^L \left[ \hat{c}_j^\dagger\hat{c}_{j+m}^\dagger + \hat{c}_j\hat{c}_{j+m} \right].
\end{align}
In a similar vein, following the earlier computation of $\mathcal{K}_{jl}$ and taking $L\gg 1$ limit, we obtain the following long-ranged real-space Hamiltonian 
\begin{align}
    \delta \hat H _y = \frac{2\gamma_y}{\pi} \sum_j \sum_{r>0, ~r\in odd} \frac{1}{r}\left[ \hat{c}_j^\dagger\hat{c}_{j+r}^\dagger + \hat{c}_j \hat{c}_{j+r} \right]
\end{align}

\subsection{Mixing $x$ and $y$ deformation: nonreciprocal Hamiltonians}
Clearly, it is evident that if we consider the deformation Hamiltonian as 
\begin{align}
    \delta\hat H_{x+y} = - i \gamma_x\sum_{k>0} F_x(k) \tilde\Psi_k^\dagger \sigma^x \tilde\Psi_k -i \gamma_y \sum_{k>0} F_y(k) \tilde\Psi_k^\dagger \sigma^y \tilde\Psi_k,
\end{align}
and choose $F_x(k) = F_y(k) = \sin{(mk)}$, then we get the following nonreciprocal Hamiltonian
\begin{align}
    \delta \hat H_{x+y} = \sum_{j=1}^L \left( \frac{\gamma_y + i \gamma_x}{2} \right) \hat{c}_j^\dagger\hat{c}_{j+m}^\dagger + \left( \frac{\gamma_y - i \gamma_x}{2} \right) \hat{c}_j\hat{c}_{j+m},
\end{align}
where non-reciprocity appears in the hopping strength.  

A similar computation for $\alpha=z$ readily follows from the above discussion.
\end{document}